\documentclass[a4paper,11pt]{article}
\pdfoutput=1 

\usepackage{jheppub} 
\usepackage{slashed}
\usepackage{braket}
\usepackage[T1]{fontenc} 
\usepackage[normalem]{ulem}
\usepackage{amsmath}      
\usepackage{array}
\title{\boldmath Krylov Complexity for Plane Wave Matrix Model}

\author[a]{Dibakar Roychowdhury}

\affiliation[a]{Department of Physics, Indian Institute of Technology Roorkee,\\Roorkee 247667, Uttarakhand, India}

\abstract{We study Krylov complexity in BMN Plane Wave Matrix Model at large mass deformation. We consider various consistent reductions of the matrix model that allow us to perform a Hamiltonian analysis which leads to different notions of the Krylov complexity. In the first part of the paper, we study the Krylov state complexity considering systematic reduction of $N=3$ and $N=4$ representations of the matrix model, which reveals a universal characteristic scaling for the Lanczos coefficients and fix them completely in terms of the mass deformation parameter. In the second part of the paper, we study the Krylov operator growth in the matrix model and compute the corresponding Lanczos coefficients. In both cases, we observe a \emph{linear} scaling of Lanczos coefficients with the mass parameter. The early time growth in Krylov complexity receives quadratic correction due to the presence of the massive deformation in the matrix model. Our analysis reveals that such massive corrections appear at same order in time for both the notion of the Krylov complexity.}
\begin{document} 
\maketitle
\flushbottom
\section{Introduction and General Idea}
In recent years, the Krylov complexity \cite{Parker:2018yvk}-\cite{Caputa:2024vrn} has emerged as one of the cornerstones of modern theoretical and computational physics that plays a seminal role in diagnosing the dynamics of chaotic systems \cite{Hashimoto:2023swv}-\cite{Bhattacharjee:2024yxj}, in understanding the holographic correspondence \cite{Caputa:2024sux}-\cite{Erdmenger:2022lov} and in the physics of black holes \cite{Kar:2021nbm}. Generally, the Krylov complexity is categorized into two classes. The Krylov state complexity (KSC) is a typical measure of the spread of the initial wavefunction (or state) in the Schrodinger description of quantum mechanics. On the other hand, the Krylov operator complexity (KOC) is a typical measure of the growth of an operator in a Krylov subspace which obeys the Heisenberg equation of motion.

In both of the above descriptions, one minimizes the cost function, where the dynamics boils down to a one dimensional lattice model. Going from one Krylov state to the next corresponds to a hopping between different lattice sites whose probability is given by $|b_n|^2$, where $b_n$ is a set of Lanczos coefficients. On the other hand, the probability of hopping between the same lattice sites is characterized by $|a_n|^2$, where $a_n$ are the remaining set of Lanczos coefficients. The Krylov complexity is uniquely fixed by these Lanczos coefficients. In the following, we carry out an \emph{analytical} calculation to show this explicitly.

The purpose of this paper is to carry out a detailed comparative analysis of the Krylov state complexity \cite{Balasubramanian:2022tpr} and the Krylov operator growth \cite{Parker:2018yvk} for the BMN Plane Wave Matrix Model (PWMM) \cite{Berenstein:2002jq}, which plays a central role in understanding the holographic correspondence and offers a controllable quantum gravity setup. We use several consistent reductions of the matrix model, which allows us to carry out a Hamiltonian analysis of the system and eventually compute and compare different notions of complexity in the matrix model. These models are broadly classified into two classes- (i) Fuzzy Sphere Model (FS) and (ii) Coulomb potential model \cite{Asano:2015eha}. The FS model has been categorized into two classes- the integrable fuzzy sphere model (IFS) and the pulsating fuzzy sphere model (PFS). 

We begin by reviewing the general setup in the matrix model and the corresponding dynamics that eventually allow us for various consistent reductions.

\subsection{BMN matrix model: short review}
The BMN matrix model \cite{Berenstein:2002jq} was initially proposed as a theory of membranes on 11d maximally super-symmetric pp wave background \cite{Dasgupta:2002hx}-\cite{Sugiyama:2002rs}. The theory has a mass gap $\mu$, which could be thought of as the massive deformation of the BFSS matrix model \cite{Banks:1996vh}.

The dual gravitational counter part has different manifestations, for example, could be thought of as an electrostatic potential problem with a set of conducting disks \cite{Lin:2004nb}-\cite{Asano:2012zt} representing the partition of $N$ in the matrix model or as model due to Polchinski and Strassler \cite{Polchinski:2000uf} where massive deformation is sourced due to background fluxes \cite{Lin:2004kw}. The geometry in all these examples is asymptote to the near horizon geometry of $N$- D0 branes.  

The BMN Plane Wave Matrix Model (PWMM) is obtained starting from $\mathcal{N}=4$ SYM on $R \times S^3$. The resulting action is obtained by modding out the fields in successive steps \cite{Asano:2012zt}. The first step is to dimension reduction along the $S^1$ fiber inside the three sphere $S^3$, which integrate out the fields that have periodicity along the $S^1$ fiber. The resulting theory is $2+1$ SYM on $R \times S^2$. Finally, the PWMM is obtained following a reduction along $S^2$ (in other words, removing the coordinate dependence along two sphere), which yields \cite{Asano:2015eha}
\begin{align}
\label{e1.1}
    S_{PWMM}&=\frac{1}{g^2}\int dt \text{Tr}\Big[ \frac{1}{2}(D_t X^r)^2+\frac{1}{4}[X^r,X^s]^2 -\frac{1}{2}\Big(\frac{\mu}{3}\Big)^2X^2_i \nonumber\\
    &-\frac{1}{2}\Big(\frac{\mu}{6}\Big)^2X^2_a -\frac{i\mu}{3}\epsilon_{ijk}X^i X^j X^k\Big]+\text{Fermions}\nonumber\\
    &\equiv \frac{1}{g^2}\int dt \text{Tr}\mathcal{Z}+\text{Fermions}.
\end{align}

Notice that in what follows, we would be concerned with the bosonic sector of the theory only. Here, $i,j=1,2,3$ are the $SU(2)$ indices, while on the other hand, $a,b=4,\cdots ,9$ are the $SO(6)$ indices. The theory has 16 supercharges, where $X^r (r=1,\cdots , 9)$ are the $U(N)$ gauge invariant matrices that transform as $X \rightarrow U X U^{-1}$. Here, $\mu$ is the mass deformation parameter that introduces a mass gap in the theory, thereby making it non-conformal. 

In the limit $\mu \rightarrow 0$, the theory approaches yet another supersymmetric theory in UV, which is known as the BFSS matrix model \cite{Banks:1996vh}. On the other hand, $\mu \neq 0$ introduces the Myers term \cite{Myers:1999ps}, which in the dual gravitational picture corresponds to the polarization of the charge of the D0 brane into the D2 branes. The gauge covariant derivative can be expressed as
\begin{align}
    D_t X^r=\partial_t X^r-i[A,X^r].
\end{align}

In what follows, we will calculate equations of motion for fields $X^i$ and $X^a$ and will figure out nontrivial solutions with $X^r \neq 0$, which are known as the \emph{fuzzy} sphere vacuum. By fuzzy sphere vacuum, we always refer to the fact that the vacuum solutions will fall under a representation of the $SU(2)$ algebra. This is precisely the case with all the examples $X^i$ and $X^a$ we consider in the subsequent analysis. At the end of our computation we set $A=0$, which corresponds to a particular choice of the (Weyl) gauge \cite{Asano:2015eha}. We express the matrices in a $N \times N$ dimensional representation as
\begin{align}
\label{e1.3}
    X^i(t)=x(t)\mathfrak{X}^i_{N \times N}~;~X^a(t)=y(t)\mathfrak{Y}^a_{N \times N}.
\end{align}

Using \eqref{e1.3}, one finds an explicit expression for $\mathcal{Z}$ up to $\mathcal{O}(A)$
\begin{align}
    \mathcal{Z}&=\frac{1}{2}(\partial_t x)^2 (\mathfrak{X}^i)^2+\frac{1}{2}(\partial_t y)^2 (\mathfrak{Y}^a)^2-\frac{i}{2}x\partial_t x \Big(\mathfrak{X}^i[A,\mathfrak{X}^i]+[A,\mathfrak{X}^i]\mathfrak{X}^i\Big)\nonumber\\
    &-\frac{i}{2}y\partial_t y \Big(\mathfrak{Y}^a[A,\mathfrak{Y}^a]+[A,\mathfrak{Y}^a]\mathfrak{Y}^a\Big)+\frac{x^4}{4}[\mathfrak{X}^i,\mathfrak{X}^j][\mathfrak{X}^i,\mathfrak{X}^j]\nonumber\\
    &+\frac{y^4}{4}[\mathfrak{Y}^a,\mathfrak{Y}^b][\mathfrak{Y}^a,\mathfrak{Y}^b]+\frac{x^2 y^2}{2}[\mathfrak{X}^i,\mathfrak{Y}^a][\mathfrak{X}^i,\mathfrak{Y}^a]-\frac{x^2 }{2}\Big(\frac{\mu}{3}\Big)^2 (\mathfrak{X}^i)^2 \nonumber\\
    &-\frac{y^2}{2}\Big(\frac{\mu}{6}\Big)^2 (\mathfrak{Y}^a)^2-\frac{i\mu}{3}x^3\epsilon_{ijk}\mathfrak{X}^i \mathfrak{X}^j \mathfrak{X}^k+\mathcal{O}(A^2).
\end{align}

Considering the variation of the classical action \eqref{e1.1}, one finds
\begin{align}
    \delta S_{PWMM}=\frac{1}{g^2}\int dt \text{Tr}\delta \mathcal{Z}
\end{align}
where the integral can be expressed as
\begin{align}
   &\text{Tr}\delta \mathcal{Z}= \text{Tr}\Bigg[-\ddot{X}^i \delta X^i-\ddot{X}^a \delta X^a+i[X^r,\dot{X}^r]\delta A+[X^r,[X^i,X^r]]\delta X^i\nonumber\\
   &+[X^r,[X^a,X^r]]\delta X^a-\Big(\frac{\mu}{3}\Big)^2 X^i \delta X^i - \Big(\frac{\mu}{6}\Big)^2 X^a \delta X^a-i \mu \epsilon_{ijk}X^j X^k \delta X^i\Bigg]+\mathcal{O}(A^2)
\end{align}
where we define $\delta X^i(t) =\delta x (t) \mathfrak{X}^i_{N \times N}$ and $\delta X^a (t)=\delta y(t)\mathfrak{Y}^a_{N \times N}$.

The equations of motion are finally obtained by setting
\begin{align}
\label{e1.7}
    &\frac{\delta S_{PWMM}}{\delta X^i}\Big|_{A=0}=0=\ddot{X}^i+[X^r,[X^r,X^i]]+\Big(\frac{\mu}{3}\Big)^2 X^i+i \mu \epsilon^{ijk}X_j X_k\\
    \label{e1.8}
    &\frac{\delta S_{PWMM}}{\delta X^a}\Big|_{A=0}=0=\ddot{X}^a+[X^r,[X^r,X^a]]+\Big(\frac{\mu}{6}\Big)^2 X^a.
\end{align}

The above equations \eqref{e1.7}-\eqref{e1.8} are supplemented by the Gauss law constraint \cite{Asano:2015eha}
\begin{align}
    &\frac{\delta S_{PWMM}}{\delta A}\Big|_{A=0}=0=[X^r,\dot{X^r}].
\end{align}

In the following, we consider various consistent reductions of the matrix model that satisfy the equations of motion \eqref{e1.7}-\eqref{e1.8}. Any non-trivial solution of the dynamics \eqref{e1.7}-\eqref{e1.8} would correspond to a fuzzy sphere vacuum of the matrix model. We explore the dynamics of these reduced systems in a Hamiltonian framework, where we create subsequent Krylov states \cite{Balasubramanian:2022tpr} under the Hamiltonian time evolution of the initial state $\ket{\Psi_0}$.
\subsection{Summary of results}
The remainder of the paper is organized as follows. The paper is largely divided into two parts. In the first part (Section \ref{sec2}), we carry out a detailed analysis of the Krylov state complexity for $N=3$ and $N=4$ representations \cite{Asano:2015eha} of the BMN Plane Wave Matrix Model \cite{Berenstein:2002jq}. We obtain the corresponding Lanczos coefficients and the associated Krylov state complexity. It appears that at large mass deformation $\mu \rightarrow \infty$, the initial wave function (which is also the Krylov state at $t=0$) corresponds to two \emph{localized} harmonic oscillators around the respective minima of their potential functions \cite{Amore:2024ihm}-\cite{Huh:2024ytz}. We consider the Hamiltonian evolution of this initial wave function (at $t=0$), which generates the subsequent Krylov states at later times and finally leads to the notion of the Krylov complexity.

\begin{table}[!h]
\begin{center}
\begin{tabular}{||c c c c||} 
 \hline
 IFS (KSC) & IFS (KOC) & PFS (KSC)  & PFS (KOC) \\ [0.5ex] 
 \hline\hline
 $a_0=1.25 \mu$ & $a_0=0$ & $a_0=1.15 \mu$ & $a_0=0$ \\ 
 \hline
 $a_1=3.75 \mu$ & $a_1=0$& $a_1=3.42 \mu$ & $a_1=0$ \\
 \hline
 $b_1=0.75 \mu$ &  $b_1=2 \mu$ & $b_1=0.84 \mu$ & $b_1=2 \mu$ \\
 \hline
 $b_2= 14 \mu$ & $b_2=11.31 \mu$ & $b_2=10.82 \mu$ & $b_2=10.80 \mu$ \\
 \hline
\end{tabular}
\caption{We summarize different Lanczos coefficients for the FS Model.}
\label{table}
\end{center}
\end{table}

Our results can be broadly summarized as follows. In both examples, we observe a \emph{universal} characteristic of the Lanczos coefficients at strong coupling, that is, $a_n = \alpha_n \mu$ and $b_n=\beta_n \mu$, where $\mu \rightarrow \infty$ is the mass deformation parameter. Here, $\alpha_n$ and $\beta_n$ are constants that depend on the particular reduction ansatz and the nature of complexity. For example, $\alpha_n=0$ for the Krylov operator complexity \cite{Parker:2018yvk}. On the other hand, the coefficients $\beta_n$ are different for the state complexity and the operator complexity. We compare different Lanczos coefficients for the Fuzzy Sphere Model (FS) \cite{Asano:2015eha} in the Krylov state (KSC) and the growth of the operator complexity (KOC) at large mass deformation in Table \ref{table}.

In the second part of the paper (Section \ref{sec3}), we perform an analog computation of the Krylov operator complexity, where we propose an initial operator $\mathcal{O}_0$, which corresponds to localized harmonic oscillator states when acts of specific reference state of the theory. The examples we study can be broadly categorized into two classes. In the first part, we consider the example of the pulsating fuzzy sphere (PFS) and the integrable fuzzy sphere (IFS) \cite{Asano:2015eha} together, while in the second part we discuss the example of the Coulomb potential. 

In either of the cases, the complexity can be schematically expressed as
\begin{align}
\label{e1.10}
    C(t)|_{t\sim 0}=\zeta \mu^2 t^2+\mathcal{O}(t^4)
\end{align}
where $\zeta\geq 0$ is a model dependent parameter that controls the early time growth of complexity. As we show, $\zeta$ can be uniquely fixed in terms of the first four Lanczos coefficients. Notice that the first non-trivial correction to complexity appears at $\mathcal{O}(\mu^2 t^2)$, which turns out to be \emph{universal} characteristic \cite{Roychowdhury:2026vzq} of the Plane Wave Matrix Model. In the above expansion \eqref{e1.10}, the entity $\mu t$ is kept fixed in the limit $\mu \rightarrow \infty$. As a comment, choosing normalization of the initial state $\ket{\Psi(0)}$ other than one would trigger a non-zero reference value for the complexity, that is, $C(0)\neq 0$, which does not alter the qualitative physics.

Finally, we draw our conclusion with some future remarks in Section \ref{sec4}.

\section{Krylov state complexity for matrix model}
\label{sec2}
The Krylov state complexity for the fuzzy sphere model \cite{Asano:2015eha} has been discussed in a recent paper \cite{Roychowdhury:2026vzq}. The purpose of this Section is to extend the analysis of \cite{Roychowdhury:2026vzq} for two more reduction ansatz, as discussed in \cite{Asano:2015eha}. These reductions follow from higher dimensional representations of $\mathfrak{X}^i$ and $\mathfrak{Y}^a$. In what follows, we are mostly concerned with the large deformation limit of the matrix model. We work out the Lanczos procedure and compute the first few Lanczos coefficients analytically, for both reduction ansatz.
\subsection{$N=3$ model: Coulomb potential}
As a first example, we consider the $3 \times 3$ representation of the matrices $\mathfrak{X}^i$ and $\mathfrak{Y}^a$, see \cite{Asano:2015eha} and \cite{Arnlind:2003nh} for details. The system is characterized by two coupled anharmonic oscillators whose dynamics can be obtained from a Lagrangian density of the form
\begin{align}
    &\mathcal{L}=\frac{1}{2}(\dot{x}^2+\dot{y}^2)-V(x,y)\\
    &V(x,y)=2 \bar{\mu}^2 \Big(x^2+\frac{y^2}{4}\Big)+2 y^4+\frac{\mathcal{Q}^2}{2y^2}+6 x^2 y^2
    \label{e2.2}
\end{align}
where $\bar{\mu}=\frac{\mu}{6}$ is a rescaled mass parameter. On the other hand, $\mathcal{Q}=y^2 \dot{\phi}$ is a conserved charge of the system \cite{Asano:2015eha}, where $\phi (t)$ is the associated cyclic coordinate. 

The charge $\mathcal{Q}$ gives rise to a repulsive Coulomb type potential. The corresponding canonical Hamiltonian is given by
\begin{align}
\label{e2.3}
    \mathcal{H}=\frac{1}{2}(p^2_x +p^2_y)+V(x,y).
\end{align}

We begin by considering the large deformation limit, that is, $\bar{\mu}\rightarrow \infty$. In order to build up the Krylov basis, we start with the minimal energy configuration where we consider the following scaling of the coordinates
\begin{align}
    x \rightarrow \lambda x~;~y \rightarrow \lambda y~;~t \rightarrow t/\lambda ~;~\bar{\mu}\rightarrow \lambda \bar{\mu}~;~\phi \rightarrow \phi~;~\mathcal{Q}\rightarrow \lambda^3 \mathcal{Q}
\end{align}
which shifts the Hamiltonian \eqref{e2.3} as $\mathcal{H}\rightarrow \lambda^4 \mathcal{H}$. Notice that $\mathcal{Q}$ scales at a faster rate than the coordinate $y$, as one approaches the origin around $y \sim 0$. In other words, in order to have the harmonic oscillator ansatz valid, one has to set $\mathcal{Q}^2/y^2\sim \lambda^4 \sim 0$, which is valid for small values of $\mathcal{Q}\ll 1$. This will be further confirmed towards the end of our calculation.

Clearly, the minimum energy configuration would correspond to setting the scaling parameter $\lambda \rightarrow 0$. In the presence of non-zero mass deformation, this would naturally correspond to setting $\bar{\mu}\rightarrow \infty$, such that $\bar{\mu}x$ and $\bar{\mu}y$ remain finite. On the other hand, the Coulomb term vanishes, since the numerator scales at a rate faster than the denominator. 

In summary, we have two localized Harmonic oscillators \cite{Amore:2024ihm}, with an effective potential
\begin{align}
    V(x,y)=2 \bar{\mu}^2 x^2 +\frac{1}{2}\bar{\mu}^2 y^2=\frac{1}{2}\bar{\mu}w^2_x x^2+\frac{1}{2}\bar{\mu}w^2_y y^2.
\end{align}
Here, $w_x=2 \sqrt{\bar{\mu}}$ and $w_y=\sqrt{\bar{\mu}}$ are fundamental frequencies of the oscillators.

We introduce the following change of variables
\begin{align}
    \hat{x}=\frac{2x}{w_x}~;~\hat{y}=\frac{y}{w_y}.
\end{align}

This yields the following kinetic energy operator (in units where we set $\hbar =c=l=1$)
\begin{align}
    &\nabla^2=-\frac{1}{2}(\partial^2_x+\partial^2_y)\rightarrow - \frac{1}{2 \bar{\mu}}\hat{\nabla}^2\\
    &\hat{\nabla}^2=\partial^2_{\hat{x}}+\partial^2_{\hat{y}}.
\end{align}

This finally yields the Schrodinger equation of the form
\begin{align}
\label{e2.9}
    -\frac{1}{2 \bar{\mu}}\hat{\nabla}^2 \Psi_0 +V(\hat{x},\hat{y})\Psi_0=E_0 \Psi_0
\end{align}
where the potential could be formally expressed as
\begin{align}
   &V(\hat{x},\hat{y})=\frac{1}{2}\bar{\mu}(\hat{w}^2_x \hat{x}^2+\hat{w}^2_y \hat{y}^2) \\
   &\hat{w}_x=\sqrt{\bar{\mu}}w_x~;~\hat{w}_y=\sqrt{\bar{\mu}}w_y.
\end{align}

In summary, modulo some coordinate redefinition, the Schordinger equation \eqref{e2.9} is the same as the one that follows from \eqref{e2.3}, which we consider as the reference Hamiltonian for constructing the Krylov basis \cite{Balasubramanian:2022tpr},\cite{Hashimoto:2023swv}. The corresponding ground state wavefunction has the standard form $\Psi_0(\hat{x},\hat{y})\sim e^{-\frac{\bar{\mu}}{2}(\hat{w}_x \hat{x}^2+\hat{w}_y \hat{y}^2)}$. We consider this wave function as the reference state on which the Hamiltonian \eqref{e2.3} acts successively to generate the Krylov subspace. Following a suitable rescaling of the coordinates, that is, $\hat{x}\rightarrow \sqrt{\hat{w}_x/2} \hat{x}$ and $\hat{y}\rightarrow \sqrt{\hat{w}_y/2} \hat{y}$, the ground state of the configuration can finally be expressed in the simple form \cite{Roychowdhury:2026vzq}
\begin{align}
\label{e2.12}
    &\ket{\Psi_0}=\int dx dy \Psi_0(x,y)\ket{x,y}\\
    & \Psi_0(x,y)=\sqrt{\frac{2 \mu}{\pi}}e^{-\mu(x^2+y^2)}
    \label{e2.13}
\end{align}
where, for simplicity, we remove the hat symbol and use $\mu$ instead of $\bar{\mu}$. 

We expand the initial state \eqref{e2.12} in the position eigen basis, such that
\begin{align}
\label{e2.14}
    \braket{x',y'|x,y}=\delta (x-x')\delta (y-y').
\end{align}

Using \eqref{e2.14}, it is straightforward to show that the initial state is normalized, that is,
\begin{align}
    \braket{\Psi_0|\Psi_0}=\int dx dy \Psi^2_0(x,y)=1.
\end{align}

Before we proceed further, the first step is to notice that a general state $\ket{\Psi(t)}$ has the Hamiltonian time evolution of the form
\begin{align}
\label{e2.16}
    \ket{\Psi(t)}=e^{-it\mathcal{H}}\ket{\Psi_0}
\end{align}
where $\ket{\Psi_0}=\ket{\Psi(t=0)}$ is the initial state.

Expanding \eqref{e2.16}, one finds the following
\begin{align}
     \ket{\Psi(t)}=\sum_{n=0}^{\infty}\frac{(-it)^n}{n!}\mathcal{H}^n \ket{\Psi_0}=\sum_{n=0}^{\infty}\frac{(-it)^n}{n!}\ket{\Psi_n}.
\end{align}

Here, we introduce a basis $\{\ket{\Psi_n}\}=\{\mathcal{H}^n\ket{\Psi_0}\}$, such that different basis elements can be obtained as a successive application of the Hamiltonian, for example,
\begin{align}
    \ket{\Psi_1}=\mathcal{H}\ket{\Psi_0}~;~\ket{\Psi_2}=\mathcal{H}^2\ket{\Psi_0},\cdots .
\end{align}

In the following, we note down a couple of these states that would be relevant for our subsequent analysis. A straightforward calculation reveals the following states
\begin{align}
\label{e2.19}
   & \ket{\Psi_1}=\int dx dy \Phi_1(x,y)\Psi_0(x,y)\ket{x,y}\nonumber\\
   & \Phi_1(x,y)=2 \mu (1-\mu(x^2+y^2))+V(x,y)\\
   \label{ee2.20}
   &\ket{\Psi_2}=\int dx dy \Phi_2(x,y)\Psi_0(x,y)\ket{x,y}\nonumber\\
   & \Phi_2(x,y)=-\frac{1}{2y^4}\mathcal{F}(x,y)+V(x,y)\Phi_1(x,y).
\end{align}

The function $\mathcal{F}(x,y)$ can be expressed as
\begin{align}
   \mathcal{F}(x,y)&= 3 \mathcal{Q}^2+y^2 \Bigg[2 \mu  \mathcal{Q}^2+24 \mu ^2 x^4 y^4+8 \mu ^2 y^8 -2 \mu  \left(3 \mu ^3+20\right) y^6\nonumber\\
   &+\left(26 \mu ^3+36\right) y^4+\mu ^2 \left(2 \mathcal{Q}^2-11\right) y^2+2x^2 \Big( \mu ^2 \mathcal{Q}^2+16 \mu ^2 y^6\nonumber\\
   &\left(4 \mu ^3+6\right) y^2-3 \mu  \left(\mu ^3+20\right) y^4\Big)\Bigg].
\end{align}
\subsubsection{Constructing the Krylov basis}
The Hamiltonian \eqref{e2.3} can be applied recursively to construct states for $n \geq 3$. In the following, we use states \eqref{e2.19} and \eqref{ee2.20} to construct the Krylov sub-space \cite{Balasubramanian:2022tpr} and the first few elements of the Krylov basis $\{\ket {K_n}\}$, such that $\braket{K_m|K_n}=\delta_{mn}$.

Given the Krylov basis, one could express the general state $\ket{\Psi(t)}$ as 
\begin{align}
\label{e2.22}
 \ket{\Psi(t)}=\sum_{n=0}^{\infty}\psi_n(t)\ket{K_n}   
\end{align}
where the coefficients $\psi_n (t)$ are fixed by the Krylov chain condition \cite{Balasubramanian:2022tpr}.\\\\
\uline{\textbf{Gram-Schmidt orthogonalization:}}\\\\
The general procedure to construct the Krylov basis is the standard one, that is, the Gram-Schmidt orthogonality, where one uses the non-orthogonal basis states $\{\ket {\Psi_n}\}$ to construct an orthogonal one, that is, the Krylov basis $\{\ket {K_n}\}$.

We expand the new basis elements in terms of the old basis elements as \cite{Roychowdhury:2026vzq}
\begin{align}
\label{e2.23}
    \ket{K_{n+1}}=\ket{\Psi_{n+1}}-c_n \ket{K_n}-d_n \ket{K_{n-1}}~;~n=0,1,2,\cdots
\end{align}
where the coefficients $c_n$ and $d_n$ are fixed from the orthonormality criteria.

To begin with, we consider $\ket{K_{0}}=\ket{\Psi_{0}}$ as the initial state, together with $\ket{K_{-1}}=0$. Clearly, the normalization is trivially satisfied for $\ket{K_{0}}$, that is, $\braket{K_0|K_0}=1$.

Setting $n=0$, we find the first relation
\begin{align}
    \ket{K_1}=\ket{\Psi_1}-c_0\ket{K_0}.
\end{align}

Taking the inner product with $\ket{K_0}$ and considering the orthogonality of the Krylov states, that is, $\braket{K_0|K_1}=0$, we fix the coefficient as
\begin{align}
\label{e2.25}
    c_0=\braket{K_0|\Psi_1}=\int dx dy \Psi_0^2 (x,y)\Phi_1(x,y).
\end{align}

A straightforward evaluation of the integral \eqref{e2.25} reveals\footnote{If we look at the integral \eqref{e2.25} closely, the entity $\Phi_1(x,y)$ \eqref{e2.19} contains the Coulomb potential $V(x,y)$. Once we evaluate the integral in \eqref{e2.25} and expand it afterward close to the origin $x \sim 0$ and $y\sim 0$, it in fact reveals an entity $-\frac{\mu  \mathcal{Q}^2 x}{\pi y}$ at LO, which is finite in the limit both $x\rightarrow 0$ and $y\rightarrow 0$. Similarly, one can argue that other integrals would yield finite answer, which is indeed the case here.}
\begin{align}
    c_0=\mu  \left(\frac{13}{8}-2 \mathcal{Q}^2\right)+\frac{3}{4 \mu ^2}.
\end{align}

Next, we consider $n=1$, which yields the following relation
\begin{align}
    \ket{K_2}=\ket{\Psi_2}-c_1\ket{K_1}-d_1 \ket{K_0}.
\end{align}

Taking the inner product with $\ket{K_0}$ on both sides and considering the orthogonal property of the Krylov state, that is $\braket{K_0|K_2}=0$, we find the following
\begin{align}
\label{e2.28}
    d_1=\braket{K_0|\Psi_2}=\int dx dy \Psi_0^2 (x,y)\Phi_2(x,y).
\end{align}

A straightforward evaluation of the integral \eqref{e2.28} reveals the following value 
\begin{align}
    d_1=\frac{1}{192 \mu ^4}\Big[\mu ^6 \left(32 \left(8 \mathcal{Q}^2-57\right) \mathcal{Q}^2+561\right)+48 \mu ^3 \left(8 \mathcal{Q}^2+3\right)+828 \Big].
\end{align}

Similarly, the coefficient $c_1$ can be obtained following an inner product with $\ket{K_1}$ and considering the orthogonality of the Krylov states, that is, $\braket{K_1|K_2}=0$, which yields 
\begin{align}
\label{e2.30}
    c_1 N_1 = \braket{K_1|\Psi_2}=\int dx dy \Psi_0^2 (x,y)\Phi_1(x,y)\Phi_2(x,y)-c_0 d_1.
\end{align}

After performing the integral in \eqref{e2.30}, one finds the following
\begin{align}
    c_1 N_1 &= \frac{32 \mu ^3 \mathcal{Q}^6}{15}-\frac{54 \mu ^3 \mathcal{Q}^4}{5}+\frac{3}{8} \left(-61 \mu ^3+\frac{34}{\mu ^3}+48\right) \mathcal{Q}^2\nonumber\\
    &+\frac{9 }{128 \mu ^6}\left(23 \mu ^9-96 \mu ^6+32 \mu ^3+884\right).
\end{align}

Notice that the states $\ket{K_n}$ are generally not normalized. For example, here we have used the fact that the norm of the state $\ket{K_1}$ is given by
\begin{align}
\label{e2.32}
    \braket{K_1|K_1}=N_1=\int dx dy \Psi_0^2 (x,y)\Phi^2_1(x,y)-c_0^2.
\end{align}

After performing the integral \eqref{e2.32}, one finds
\begin{align}
    N_1=\frac{1}{96 \mu ^4}\Big[\mu ^6 \left(27-32 \mathcal{Q}^2 \left(8 \mathcal{Q}^2+9\right)\right)+6 \mu ^3 \left(80 \mathcal{Q}^2-27\right)+360 \Big].
\end{align}

On a similar note, the norm of $\ket{K_2}$ is
\begin{align}
    \braket{K_2|K_2}=N_2=\int dx dy \Psi_0^2 (x,y)\Phi^2_2(x,y)-N_1 c_1^2 -d_1^2.
\end{align}

After a straightforward computation, one finds the following
\begin{align}
    &N_2=-\frac{1}{36864 \mu ^8}\left(\mu ^6 \left(32 \left(8 \mathcal{Q}^2-57\right) \mathcal{Q}^2+561\right)+48 \mu ^3 \left(8 \mathcal{Q}^2+3\right)+828\right)^2\nonumber\\
    &+\frac{1}{430080 \mu ^8}\Bigg[ 2520 \mu ^6 \left(32 \left(72 \mathcal{Q}^2-95\right) \mathcal{Q}^2+2813\right)+302400 \mu ^3 \left(96 \mathcal{Q}^2+335\right)\nonumber\\
    &+814438800 -1680 \mu ^9 \left(256 \mathcal{Q}^4+6368 \mathcal{Q}^2+6033\right)+\mu^{12}\Big(7875105 +64 \mathcal{Q}^2\nonumber\\
    & \times \left(8 \mathcal{Q}^2 \left(128 \mathcal{Q}^4-4896 \mathcal{Q}^2+60083\right)-1663305\right.\Big)\Bigg]-\frac{96 \mu ^4 c^2_1 N_1^2}{\mathcal{B}}\nonumber\\
    &\mathcal{B}=\mu ^6 \left(27-32 \mathcal{Q}^2 \left(8 \mathcal{Q}^2+9\right)\right)+6 \mu ^3 \left(80 \mathcal{Q}^2-27\right)+360.
\end{align}

In summary, we have the following (normalized) Krylov states
\begin{align}
\label{e2.36}
  &\ket{K_0}=\ket{\Psi_0}\\
  &\ket{K_1}=\frac{1}{\sqrt{N_1}}\Big[ \ket{\Psi_1}-c_0\ket{K_0}\Big]\\
  &\ket{K_2}=\frac{1}{\sqrt{N_2}}\Big[ \ket{\Psi_2}-c_1\ket{K_1}-d_1 \ket{K_0}\Big].
  \label{e2.38}
\end{align}
The above construction can continue for states $n\geq 3$.
\subsubsection{Lanczos algorithm and Krylov complexity}
Given the Krylov basis \eqref{e2.36}-\eqref{e2.38}, we apply the Lanczos algorithm to find the Lanczos coefficients. The Krylov basis satisfies the Krylov chain condition \cite{Balasubramanian:2022tpr}
\begin{align}
    \mathcal{H} \ket{K_n}=a_n \ket{K_n}+b_n \ket{K_{n-1}}+b_{n+1}\ket{K_{n+1}}
\end{align}
where the Lanczos coefficients are given by
\begin{align}
\label{e2.40}
    a_n = \braket{K_n|\mathcal{H}|K_n}~;~b_n = \braket{K_{n-1}|\mathcal{H}|K_n}.
\end{align}

In the following, we compute the first few of them using the basis \eqref{e2.36}-\eqref{e2.38}. For example, for $n=0$ one finds that
\begin{align}
    a_0 (\mu , \mathcal{Q}) = \braket{K_0|\mathcal{H}|K_0}=c_0=\mu  \left(\frac{13}{8}-2 \mathcal{Q}^2\right)+\frac{3}{4 \mu ^2}.
\end{align}

Next, we set $n=1$, which in the large $\mu$ limit yields
\begin{align}
\label{e2.42}
    a_1(\mu , \mathcal{Q}) &= \braket{K_1|\mathcal{H}|K_1}\nonumber\\
    &=\frac{1}{N_1}\Big[ \braket{\Psi_1|\Psi_2}-c_0 \braket{\Psi_1|\Psi_1}-c_0 d_1+c^3_0\Big].
\end{align}

Computing each of the above terms in \eqref{e2.42}, one finds the following
\begin{align}
  &a_1(\mu , \mathcal{Q})= \frac{\mathcal{P}_1}{\mathcal{P}_2}\nonumber\\
  &\mathcal{P}_1=1980 \mu ^3 \left(5-32 \mathcal{Q}^2\right)-120 \mu ^6 \left(384 \mathcal{Q}^4+280 \mathcal{Q}^2-135\right)-227880\nonumber\\
  &+\mu ^9 \left(16 \mathcal{Q}^2 \left(768 \mathcal{Q}^4+2992 \mathcal{Q}^2+4185\right)-4455\right)\nonumber\\
  &\mathcal{P}_2=40 \mu ^2 \left(\mu ^6 \left(32 \mathcal{Q}^2 \left(8 \mathcal{Q}^2+9\right)-27\right)+6 \mu ^3 \left(27-80 \mathcal{Q}^2\right)-360\right).
\end{align}

Taking a large $\mu$ expansion, one can further simplify the expression as
\begin{align}
    &a_1(\mu , \mathcal{Q})=\mu  \left(\frac{6 \mathcal{Q}^2}{5}+\frac{36 \left(104 \mathcal{Q}^2-3\right)}{5 \left(32 \mathcal{Q}^2 \left(8 \mathcal{Q}^2+9\right)-27\right)}+\frac{133}{40}\right)\nonumber\\
    &+\frac{142155-576 \mathcal{Q}^2 \left(8 \left(640 \mathcal{Q}^4+96 \mathcal{Q}^2-2181\right) \mathcal{Q}^2+6435\right)}{20 \mu ^2 \left(27-32 \mathcal{Q}^2 \left(8 \mathcal{Q}^2+9\right)\right)^2}+\mathcal{O}(1/\mu^3).
\end{align}

Next, we note off diagonal elements $b_n$. To begin with, we notice that $b_0=0$. Next, we set $n=1$, which yields the following
\begin{align}
    b_1 (\mu, \mathcal{Q})=\braket{K_0|\mathcal{H}|K_1}=\frac{1}{\sqrt{N_1}}(d_1-c_0^2).
\end{align}

A straightforward computation of the above entities reveal
\begin{align}
   &b_1 (\mu, \mathcal{Q})= \frac{1}{4 \mu^2}\sqrt{\frac{1}{6} \mu ^6 \left(27-32 \mathcal{Q}^2 \left(8 \mathcal{Q}^2+9\right)\right)+\mu ^3 \left(80 \mathcal{Q}^2-27\right)+60} \nonumber\\
   &=\frac{\mu }{4 \sqrt{6}}\sqrt{ \left(27-32 \mathcal{Q}^2 \left(8 \mathcal{Q}^2+9\right)\right)}+\mathcal{O}(1/\mu^3).
\end{align}

Next, we set $n=2$ and the corresponding Lanczos coefficient is given by
\begin{align}
\label{e2.47}
    &b_2 (\mu, \mathcal{Q})= \braket{K_1 |\mathcal{H}|K_2}\nonumber\\
    &=\frac{1}{\sqrt{N_1 N_2}}\Big[ \braket{\Psi_1|\Psi_3}-d_1 \braket{\Psi_1|\Psi_1} -c_0 \braket{\Psi_0|\Psi_3}\nonumber\\
    &+ c^2_0 d_1\Big]-\frac{1}{N_1 \sqrt{N_2}}\Big[c_1\braket{\Psi_1|\Psi_2} -c_0 c_1 \braket{\Psi_1|\Psi_1}\nonumber\\
    &-c_0 c_1 d_1 +c^3_0 c_1\Big].
\end{align}

The state $\ket{\Psi_3}$ is given by the following expression
\begin{align}
\label{e2.48}
    \ket{\Psi_3}&=\mathcal{H}^3 \ket{\Psi_0}\nonumber\\
    &=\int dx dy \Phi_3 (x,y)\Psi_0(x,y)\ket{x,y}\\
     \Phi_3(x,y)&=-\frac{1}{2y^6}\mathcal{G}(x,y)+V(x,y)\Phi_2(x,y).
\end{align}

The detailed expression of the function $\mathcal{G}(x,y)$ is given by
\begin{align}
    &\mathcal{G}(x,y)=5 \mathcal{Q}^2 \left(\mathcal{Q}^2-6\right)+16 \mu ^2 y^{16}+\mu ^2 \mathcal{Q}^2 \left(\mathcal{Q}^2-6\right) y^4+\mu  \mathcal{Q}^2 \left(\mathcal{Q}^2-6\right) y^2 \left(\mu  x^2+3\right)\nonumber\\
    &+2 y^6 \left(\mu ^3 \left(3 \mathcal{Q}^2-23\right)+8 \left(\mathcal{Q}^2-3\right)+12 \mu ^2 \left(\mathcal{Q}^2-1\right) x^4+\mu ^4 \left(11-3 \mathcal{Q}^2\right) x^2-36 \mu  \left(\mathcal{Q}^2-3\right) x^2\right)\nonumber\\
    &-8 \mu  y^{14} \left(3 \left(\mu ^3+6\right)-14 \mu  x^2\right)+y^{12} \left(9 \mu ^6+264 \mu ^3+240 \mu ^2 x^4-96 \left(\mu ^3+9\right) \mu  x^2+272\right)\nonumber\\
    &+y^{10} \left(-93 \mu ^5+8 \mu ^2 \left(\mathcal{Q}^2-96\right)+144 \mu ^2 x^6-72 \mu  \left(\mu ^3+18\right) x^4+3 \left(3 \mu ^6+296 \mu ^3+384\right) x^2\right)\nonumber\\
    &+y^8 \left(\mu  \left(193 \mu ^3-6 \left(\mu ^3+4\right) \mathcal{Q}^2+648\right)+144 \left(2 \mu ^3+3\right) x^4-8 \mu ^2 x^2 \left(6 \mu ^3-4 \mathcal{Q}^2+219\right)\right).
\end{align}

Using \eqref{e2.48}, we finally arrive at the following expression
\begin{align}
  b_2 (\mu, \mathcal{Q})=\frac{271 \mu }{8 \sqrt{3}}-\frac{253}{3 \sqrt{6}}+\mathcal{O}(1/\mu) 
\end{align}
which is subjected to an expansion in small $\mathcal{Q}$ followed by an expansion in large $\mu$.

Clearly, at LO in the large $\mu$ expansion, we have a generic structure for the Lanczos coefficients, that is, $a_n = \alpha_n \mu$ and $b_n=\beta_n \mu$. This is in accordance with the previous observation in the context of the fuzzy sphere model \cite{Roychowdhury:2026vzq}, and appears to be a generic feature of the Krylov state complexity in the BMN matrix model.

With the above coefficients estimated, we are now in a position to calculate the Krylov complexity for $N=3$ BMN matrix model. The Krylov state complexity is defined as \cite{Balasubramanian:2022tpr}
\begin{align}
    C(t)=\sum_{n=0}^\infty n | \psi_n(t)|^2
\end{align}
where $\psi_n(t)$ are the expansion coefficients as defined in \eqref{e2.22}.

Considering the Schrodinger equation
\begin{align}
    i \partial_t \ket{\Psi (t)}=i\sum_{n=0}^\infty \partial_t \psi_n(t)\ket{K_n}=\mathcal{H}\ket{\Psi (t)}=\sum_{n=0}^\infty  \psi_n(t)\mathcal{H}\ket{K_n}
\end{align}
and thereby using the Krylov chain condition \eqref{e2.23}, one finally obtains
\begin{align}
     i \partial_t \psi_n (t)=a_n \psi_n (t)+b_{n+1}\psi_{n+1}(t)+b_n \psi_{n-1}(t).
\end{align}

Setting $n=0$, we find the first equation connecting the coefficients $\psi_0 (t)$ and $\psi_1 (t)$
\begin{align}
    i \partial_t \psi_0 (t) = a_0 \psi_0(t)+b_1 \psi_1 (t).
\end{align}

We calculate $\psi_1(t)$ by equating the following expansions
\begin{align}
     & \ket{\Psi(t)}=\ket{\Psi_0}-it \ket{\Psi_1}+\cdots\\
    & \ket{\Psi(t)}=\psi_0(t)\ket{K_0}+\psi_1(t)\ket{K_1}+\cdots 
\end{align}
and thereby taking the inner product with $\ket{K_1}$, which yields
\begin{align}
\label{e2.58}
    \psi_1 (t)=-it \braket{K_1|\Psi_1}=-i t \braket{K_1|\mathcal{H}|K_0}=-itb_1.
\end{align}

This yields the zeroth order coefficient $\psi_0(t)$ as
\begin{align}
 \psi_0(t)= \frac{b_1^2 }{a^2_0}(-1+i a_0 t)+c_1 e^{-i a_0 t}  
\end{align}
where $c_1$ is the integration constant. 

As we elaborate below, the choice of the constant $c_1$ is merely shifting the reference of the complexity (at $t=0$) from zero to non-zero value, which does not alter the qualitative physics much. Notice that if we set $c_1=1$ then at $t=0$, $\psi_0(0)=1-\frac{b_1^2}{a^2_0}$, which is a constant and does not depend on the mass parameter $\mu$. This would correspond to an initial state $\ket{\Psi(t)}$ that is \emph{not} normalized to one. However, we can normalize it to one by choosing the integration constant $c_1=1+\frac{b^2_1}{a^2_0}$. In what follows, we choose to work with this choice of $c_1$ that sets $\psi_0(0)=1$ and hence $\braket{\Psi(0)|\Psi(0)}=1$ \cite{Hashimoto:2023swv}.

Finally, for $n=1$, we have the following equation for the wave function
\begin{align}
     i \partial_t \psi_1 (t)=a_1 \psi_1(t)+b_2 \psi_2(t)+b_1 \psi_0(t).
\end{align}

\begin{figure}
    \centering
    \includegraphics[width=0.6\linewidth]{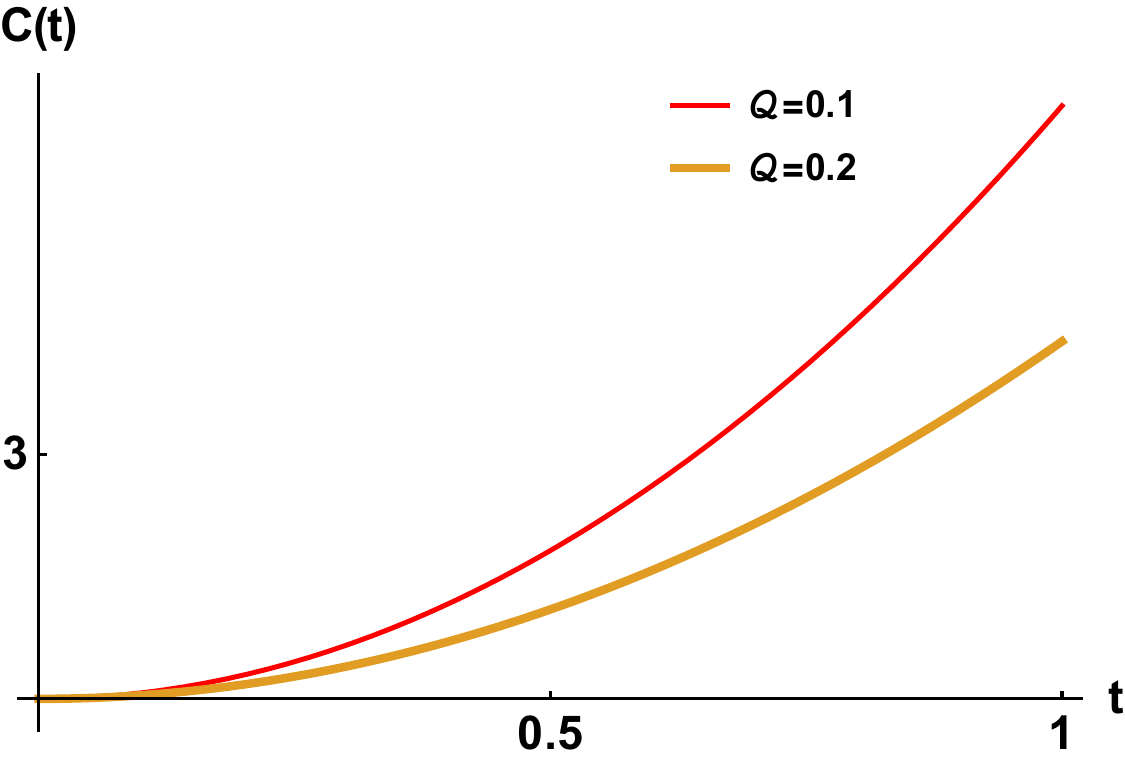}
    \caption{We plot early time growth in complexity for different values of the Coulmb charge $\mathcal{Q}$.}
    \label{fig1}
\end{figure}

After some algebra, one finally obtains
\begin{align}
\label{e2.61}
    \psi_2(t)=\frac{b_1}{b_2}-\frac{a_1}{b_2}\psi_1(t)-\frac{b_1}{b_2}\psi_0(t).
\end{align}

Notice that $\psi_1(0)=0$. On the other hand, with our choice of normalization $c_1=1+\frac{b^2_1}{a^2_0}$, we fix $\psi_0(0)=1$. In other words, this will automatically set $\psi_2(0)=0$. Combining these facts, one can set $C(0)=0$. In other words, $C(0)\neq 0$ is merely a choice of $\psi_0(0)$.

Using \eqref{e2.58} and \eqref{e2.61}, the Krylov complexity turns out to be
\begin{align}
\label{e2.62}
     C(t)|_{t \sim 0}&=| \psi_1(t)|^2+2| \psi_2(t)|^2 + \cdots \nonumber\\
    &=\zeta \mu^2 t^2 +\mathcal{O}(t^4).
\end{align}

The coefficients in \eqref{e2.62}, can be expressed (in small $\mathcal{Q}$ and large $\mu$ expansion) as
\begin{align}
\label{e2.63}
    &\zeta (\mathcal{Q})=\frac{b^2_1}{\mu^2}\Big[1+\frac{2}{ b_2^2}\left(a_0 +a_1\right)^2\Big]=0.329871-3.85685 \mathcal{Q}^2 +\mathcal{O}(\mathcal{Q}^4).
\end{align}

Clearly, the first nontrivial correction to complexity appears at order $\mu^2$, subjected to the fact that the mass deformation $\mu$ is large. A similar feature has been observed for the fuzzy sphere model \cite{Roychowdhury:2026vzq}, which therefore appears to be the universal characteristic of the BMN matrix model at strong coupling. However, contrary to the previous analysis \cite{Roychowdhury:2026vzq}, we have an additional dependence of the complexity on the global Coulomb charge $\mathcal{Q}$ of the system. Taking into account that $\zeta \geq 0$, we have a bound $\mathcal{Q}_C=0.292453$. In Fig.\ref{fig1} we plot the complexity for $\mu=50$ and different choices of $\mathcal{Q}<\mathcal{Q}_C$, which exhibit quadratic growth at early time scales. Fig.\ref{fig1} clearly reveals that the complexity grows at faster rate for smaller values of the Coulomb charge $\mathcal{Q}$, which is also evident from \eqref{e2.63}. The upper limit of $\mathcal{Q}<1$, is an artifact of the finiteness of the potential near the origin. It simply reflects to the fact that complexity is not well defined for larger values of the Coulomb charge $\mathcal{Q}$. 
\subsection{$N=4$ model: Integrable fuzzy sphere}
The model corresponds to $4 \times 4$ representation of the matrices \cite{Asano:2015eha}
\begin{align}
 X^i= \frac{1}{2} \begin{pmatrix}
y(t)\sigma^i & x(t)\sigma^i \\
x(t)\sigma^i & y(t)\sigma^i
\end{pmatrix} ~;~X^a=0
\end{align}
where $\sigma^i$ are Pauli matrices. The model corresponds to a system of two independent oscillators whose integrability was shown in \cite{saha}.

The corresponding Lagrangian can be expressed as
\begin{align}
    &L=\frac{1}{2}(\dot{x}^2+\dot{y}^2)-V(x,y)\\
    &V(x,y)=\frac{\mu^2}{2}(x^2+y^2)+\frac{1}{2}(x^4+y^4)+3x^2 y^2 -\mu y^3-3 \mu x^2 y.
\end{align}

The phase space is 4 dimensional and the system is completely integrable. Therefore, there exist two conserved charges. The trivial one is the Hamiltonian
\begin{align}
    \mathcal{H}=\frac{1}{2}(p_x^2+p_y^2)+V(x,y).
\end{align}

Like before, one can consider a scaling of the form
\begin{align}
     x \rightarrow \lambda x~;~y \rightarrow \lambda y~;~t \rightarrow t/\lambda ~;~\mu\rightarrow \lambda \mu
\end{align}
and thereby consider the limit $\lambda \rightarrow 0$. This would lead to the minimum energy configuration with two localized oscillators \cite{Amore:2024ihm} with the potential function
\begin{align}
    V(x,y)=\frac{\mu^2}{2}(x^2+y^2)
\end{align}
where $\mu x$ and $\mu y$ are kept finite in the limit $x\rightarrow 0$ and $y\rightarrow 0$.
\subsubsection{Constructing the Krylov basis}
Like before, we choose the initial state as in \eqref{e2.12}. The subsequent states are given by
\begin{align}
    & \ket{\Psi_1}=\int dx dy \Phi_1(x,y)\Psi_0(x,y)\ket{x,y}\nonumber\\
   & \Phi_1(x,y)=2 \mu (1-\mu(x^2+y^2))+V(x,y)\\
   \label{e2.20}
   &\ket{\Psi_2}=\int dx dy \Phi_2(x,y)\Psi_0(x,y)\ket{x,y}\nonumber\\
   & \Phi_2(x,y)=\mathcal{F}(x,y)+V(x,y)\Phi_1(x,y)
\end{align}
where the function $\mathcal{F}(x,y)$ is given by
\begin{align}
    &\mathcal{F}(x,y)=\mu  \left(5 \left(x^4+6 x^2 y^2+y^4\right)+6 y\right)-6 \left(x^2+y^2\right)-\mu^2\Big(x^6+7 x^4 y^2-7 \nonumber\\
    &+x^2 \left(7 y^3+24\right) y+y^6+8 y^3\Big)+3 \mu ^4 \left(x^2+y^2\right)^2\nonumber\\
    &+\mu ^3 \left(x^2+y^2\right) \left(6 x^2 y+2 y^3-13\right).
\end{align}

The Krylov state at $t=0$ is defined as before, that is, $\ket{K_0}=\ket{\Psi_0}$, such that $\braket{K_0|K_0}=1$. Next, we introduce the Krylov state 
\begin{align}
    \ket{K_1}=\ket{\Psi_1}-c_0\ket{K_0}
\end{align}
where the coefficient $c_0$ is given by
\begin{align}
    c_0= \frac{5 \mu }{4}+\frac{3}{8 \mu ^2}.
\end{align}

The next Krylov state is defined as
\begin{align}
    \ket{K_2}=\ket{\Psi_2}-c_1 \ket{K_1}-d_1 \ket{K_0}.
\end{align}

The coefficients $d_1$ and $c_1$ are given by
\begin{align}
    &d_1=\frac{1}{64 \mu ^4}(136 \mu ^6+48 \mu ^3+57)\\
    &c_1N_1=\frac{3 }{64 \mu ^6}\left(60 \mu ^9-79 \mu ^6+198 \mu ^3+111\right).
\end{align}

The constants $N_1$ and $N_2$ are fixed by the normalization and are given by
\begin{align}
    &N_1=\frac{3}{16 \mu ^4} \left(3 \mu ^6-\mu ^3+4\right)\\
    &N_2=\frac{1}{256 \mu ^8 \left(3 \mu ^6-\mu ^3+4\right)}\Big[30501 +3\mu^3\Big(62606 \mu ^6-19329 \mu ^3+17706\nonumber\\
    &+6 \mu ^9 \left(54 \mu ^6+706 \mu ^3-3623\right)\Big)\Big].
\end{align}
\subsubsection{Lanczos algorithm and Krylov complexity}
We have normalized Krylov states as in \eqref{e2.36}-\eqref{e2.38}, which we further use to calculate the Lanczos coefficients \eqref{e2.40}. We first calculate the diagonal elements
\begin{align}
    a_0(\mu)=c_0=\frac{5 \mu }{4}+\frac{3}{8 \mu ^2}.
\end{align}

Next, we set $n=1$, which yields the following coefficient
\begin{align}
    a_1(\mu)=\frac{90 \mu ^9-157 \mu ^6+359 \mu ^3+210}{24 \mu ^8-8 \mu ^5+32 \mu ^2}=\frac{15 \mu }{4}-\frac{127}{24 \mu ^2}+\mathcal{O}(1/\mu^3).
\end{align}

The first off-diagonal element turns out to be
\begin{align}
    b_1(\mu)=\frac{1}{4 \mu ^2} \sqrt{9 \mu ^6-3 \mu ^3+12}=\frac{3 \mu }{4}-\frac{1}{8 \mu ^2}+\mathcal{O}(1/\mu^3).
\end{align}

Finally, we obtain the state
\begin{align}
\label{e2.86}
    \ket{\Psi_3}&=\mathcal{H}^3 \ket{\Psi_0}\nonumber\\
    &=\int dx dy \Phi_3 (x,y)\Psi_0(x,y)\ket{x,y}\\
     \Phi_3(x,y)&=\frac{1}{2}\mathcal{G}(x,y)+V(x,y)\Phi_2(x,y).
\end{align}

The function $\mathcal{G}(x,y)$ can be expressed as
\begin{align}
    &\mathcal{G}(x,y)=24-4 \left(x^2+y^2\right) \left(5 x^4+46 x^2 y^2+5 y^4\right)+3 \mu \Big[ 3 x^8+36 x^6 y^2\nonumber\\
    &+2 x^4 y \left(57 y^3+34\right)+4 x^2 \left(9 y^6+34 y^3-18\right)+y^2 \left(3 y^6+20 y^3-72\right)\Big]\nonumber\\
    &-\mu^2 \Big[x^{10}+13 x^8 y^2+x^6 \left(50 y^4+96 y\right)+x^2 y^2 \left(13 y^6+288 y^3-840\right) \nonumber\\
    &+2 x^4 \left(25 y^6+304 y^3-96\right)+y \left(y^9+32 y^6-168 y^3-168\right)\Big]\nonumber\\
    &+2 \mu^3 \Big[ 6 x^8 y+11 x^6 \left(4 y^3-3\right)+2 y^9-19 y^6-134 y^3+38 \nonumber\\
    &+7 x^4 y^2 \left(8 y^3-15\right)+x^2 y \left(20 y^6-147 y^3-342\right)\Big]+2 \mu^4 \left(x^2+y^2\right)\nonumber\\
    & \times \Big[ 3 x^6+3 x^4 y^2+3 x^2 y \left(3 y^3+56\right)+y^6+56 y^3-104\Big]-9 \mu ^6 \left(x^2+y^2\right)^3\nonumber\\
    &+3 \mu ^5 \left(x^2+y^2\right)^2 \left(12 x^2 y+4 y^3-31\right).
\end{align}

Using \eqref{e2.86}, the Lanczos coefficient \eqref{e2.47} turns out to be
\begin{align}
    b_2(\mu)=14 \mu-\frac{50}{3}+\mathcal{O}(1/\mu).
\end{align}

The Krylov (state) complexity \eqref{e2.62} exhibits a quadratic growth in time
\begin{align}
    C(t)|_{t \sim 0}= \zeta \mu^2 t^2+\mathcal{O}(t^4).
\end{align}

The constant pre-factor is given by 
\begin{align}
    \zeta=\frac{b^2_1}{\mu^2}\Big[1+\frac{2}{ b_2^2}\left(a_0 +a_1\right)^2\Big]=0.705995 +\mathcal{O}(1/\mu).
\end{align}
\section{Krylov operator complexity in matrix model}
\label{sec3}
Let us begin with some general comments on the Krylov operator growth \cite{Parker:2018yvk} in quantum mechanics and outline the algorithm that will be followed in the subsequent analysis. The Krylov operator complexity corresponds to the operator growth $\mathcal{O}(t)=e^{i\mathcal{H}t}\mathcal{O}(0)e^{-i \mathcal{H} t}$ in quantum mechanics, along the lines of the Heisenberg equation of motion \cite{Parker:2018yvk},\cite{Hashimoto:2023swv}. In the dual gravity description, this would correspond to a particle in the bulk \cite{Caputa:2024sux} which corresponds to a local unitary operator $\mathcal{O}_0=\mathcal{O}(0)$ inserted at $t=0$ in the matrix model. In the (Krylov) operator formalism of complexity, we map the local unitary operator $\mathcal{O}$ to the (initial) state $\mathcal{O}_0 \rightarrow\mathcal{O}_0\ket{ref}:=|\mathcal{O}_0)$ in the ``operator Hilbert space'' \cite{Parker:2018yvk}. Here, $\ket{ref}$ is some reference state of the theory, which in our example will serve as the position basis $\ket{x,y}$, while the operator $\mathcal{O}(t)$ will be expressed in the position representation.

The other states $|\mathcal{O}_n)$ in the operator Hilbert space (at later times) are constructed by successive application of the \emph{Liouvillian} operator $\hat{\mathcal{L}}:=[\mathcal{H},]$, such that $\hat{\mathcal{L}}^n \mathcal{O}_0 \ket{ref}:=|\mathcal{O}_n)$. By Krylov operator complexity, we refer to the spread of the operator $\mathcal{O}(t)$ (in operator Hilbert space) in a Krylov basis of the form \cite{Parker:2018yvk}, \cite{Hashimoto:2023swv}
\begin{align}
\label{e3.1}
    \mathcal{O}(t)=\sum_{n=0}^\infty \frac{(it)^n}{n!}\hat{\mathcal{L}}^n\mathcal{O}_0=\sum_{n=0}^\infty \frac{(it)^n}{n!}\mathcal{O}_n
\end{align}
where we define the $n$th basis element under time evolution as, $\mathcal{O}_n=\hat{\mathcal{L}}^n\mathcal{O}_0$.

In summary, we construct a basis $\{\mathcal{O}_n\}= \{\hat{\mathcal{L}}^n\mathcal{O}_0\}$, that is,
\begin{align}
    \mathcal{O}_1=\hat{\mathcal{L}}\mathcal{O}_0 = [\mathcal{H},\mathcal{O}_0]~;~ \mathcal{O}_2=\hat{\mathcal{L}}^2\mathcal{O}_0 = [\mathcal{H},\mathcal{O}_1]~;~\cdots.
\end{align}

The initial choice (at $t=0$) of operator is motivated by the fact that at strong mass deformation $\mu \rightarrow \infty$, the BMN Plane Wave Matrix Model is characterized by a system of localized harmonic oscillators, around their respective minima of potential \cite{Amore:2024ihm}-\cite{Huh:2024ytz}. This motivates us to propose the following operator as our initial state
\begin{align}
\label{e3.3}
    \mathcal{O}_0=\sqrt{\frac{2 \mu}{\pi}}e^{-\mu(\hat{x}^2+\hat{y}^2)}.
\end{align}

Given the position basis, $\ket{x,y}$ the above choice \eqref{e3.3} implies that
\begin{align}
    \mathcal{O}_0 \ket{x,y}=\Psi_0(x,y)\ket{x,y}
\end{align}
where $\Psi_0(x,y)$ is given in \eqref{e2.13}.

The inner product between the operators $\mathcal{O}_m$ and $\mathcal{O}_n$, which is given by
\begin{align}
\label{e3.5}
    (\mathcal{O}_n|\mathcal{O}_m)=\text{Tr}(\mathcal{O}^\dagger_n \mathcal{O}_m)=\int dx dy \bra{x,y}\mathcal{O}^\dagger_n \mathcal{O}_m \ket{x,y}.
\end{align}

Using the above definition \eqref{e3.5}, it is trivial to check that $|\mathcal{O}_0)$ is normalized
\begin{align}
    (\mathcal{O}_0|\mathcal{O}_0)=\text{Tr}(\mathcal{O}_0^\dagger \mathcal{O}_0)=\frac{2 \mu}{\pi}\int dx dy ~e^{-2\mu(x^2+y^2)}=1.
\end{align}
\subsection{Fuzzy sphere model}
The first example we consider is that of the \emph{fuzzy} sphere model \cite{Asano:2015eha}, which is given by the Hamiltonian (operator) of the following form
\begin{align}
    \mathcal{H}=\frac{1}{2}(p_x^2+p_y^2)+V(x,y).
\end{align}

The potential function can be expressed as
\begin{align}
    V(x,y)=\frac{\mu^2}{2}(k x^2+y^2)+\frac{1}{2}(x^4+y^4)+l x^2 y^2-\mu y^3-n \mu x^2 y
\end{align}
where $k=\frac{1}{4}$, $l=1$ and $n=0$ for pulsating fuzzy sphere (PFS) \cite{Asano:2015eha}. On the other hand, $k=1$, $l=3$, $n=3$ for the integrable fuzzy sphere model (IFS). In the following, we use $[x,p_x]=i=[y,p_y]$ with $p_i=-i \partial_i$ together with $\hbar=1$. For simplicity, throughout the analysis, we have removed the ``hat'' symbol for operators. We must keep in mind that $x_i$ and $p_i$ are operators in the position representation such that $x \ket{x,y}=x\ket{x,y}$.

Next, we compute the operators for $n=1$ and $n=2$. For $n=1$, we notice that
\begin{align}
    \mathcal{O}_1=[\mathcal{H},\mathcal{O}_0] =\frac{1}{2}[p_x^2+p_y^2,\mathcal{O}_0]= \frac{2i \sqrt{2}\mu^{3/2}}{\sqrt{\pi}}(p_x x+p_y y)e^{-\mu (x^2+y^2)}.
\end{align}

Next, we consider $n=2$, which reveals the following
\begin{align}
    \mathcal{O}_2=[\mathcal{H},\mathcal{O}_1]=\frac{1}{2}[p_x^2+p_y^2,\mathcal{O}_1]+[V(x,y),\mathcal{O}_1].
\end{align}

After some algebra, we find the following expression
\begin{align}
  &\mathcal{O}_2=  \frac{2 \sqrt{2}\mu^{3/2}}{\sqrt{\pi}}\Big[p^2_x+p^2_y-2 \mu (p_x x+p_y y)^2\Big]e^{-\mu (x^2+y^2)}-\frac{2 \sqrt{2}\mu^{7/2}}{\sqrt{\pi}}(kx^2+y^2)e^{-\mu (x^2+y^2)}\nonumber\\
  &-\frac{4 \sqrt{2}\mu^{3/2}}{\sqrt{\pi}}(x^4+y^4)e^{-\mu (x^2+y^2)}- \frac{8\sqrt{2}l\mu^{3/2}}{\sqrt{\pi}}x^2 y^2e^{-\mu (x^2+y^2)}\nonumber\\
  &+\frac{6\sqrt{2}\mu^{5/2}}{\sqrt{\pi}}y^3e^{-\mu (x^2+y^2)}+\frac{6\sqrt{2}n\mu^{5/2}}{\sqrt{\pi}}x^2 y e^{-\mu (x^2+y^2)}.
\end{align}

\subsubsection{Constructing the Krylov basis}
Like in the case of the Krylov state complexity, one has to construct an orthonormal Krylov basis corresponding to operator growth such that $(K_m|K_n)=\delta_{mn}$. The operator growth in the Krylov basis is given by \cite{Parker:2018yvk}, \cite{Hashimoto:2023swv}
\begin{align}
\label{e3.12}
    \mathcal{O}(t)=\sum_n i^n \varphi_n(t)K_n
\end{align}
where $\varphi_n(t)$ is fixed from the Krylov chain condition \cite{Parker:2018yvk}, which we discuss below.\\\\
\uline{\textbf{Gram-Schmidt orthogonalization:}}\\\\
The Krylov operators (or states) $\{K_n\}$ are constructed using the previously introduced operators $\{\mathcal{O}_n\}$. The following construction ensures that the diagonal entries of the Liouvillian operator ($\hat{\mathcal{L}}$) are all zero \cite{Parker:2018yvk}. Like before, this is achieved following the Gram-Schmidt orthogonality procedure. We define the following linear map (for $n \geq 0$)
\begin{align}
K_{n+1}=\mathcal{O}_{n+1}-c_{n}K_{n}-d_n K_{n-1}.
\end{align}

The initial Krylov state we choose is $K_0=\mathcal{O}_0$, which is normalized by construction, that is, $(K_0|K_0)=1$. The next Krylov state corresponds to setting $n=0$, which yields
\begin{align}
    K_1=\mathcal{O}_1-c_0 K_0
\end{align}
which is subject to the fact $K_{-1}=0$.

Taking the inner product with $|K_0)$ and setting $(K_0|K_1)=0$, we obtain the coefficient
\begin{align}
    c_0=(K_0|\mathcal{O}_1) = \text{Tr}(\mathcal{O}^\dagger_0 \mathcal{O}_1)=\int dx dy \bra{x,y}\mathcal{O}^\dagger_0 \mathcal{O}_1 \ket{x,y}.
\end{align}

After a straightforward calculation, one finds that
\begin{align}
\label{e3.15}
    &\mathcal{O}_1 \ket{x,y}=g(x,y)\ket{x,y}\\
    &g(x,y)= \sqrt{\frac{32}{\pi }} \mu ^{3/2} e^{-\mu  \left(x^2+y^2\right)} \left(1-\mu  \left(x^2+y^2\right)\right).
\end{align}

On a similar note, we find
\begin{align}
    &\mathcal{O}_2 \ket{x,y}=h(x,y)\ket{x,y}\\
    &h(x,y)=8\sqrt{\frac{2}{\pi }} \mu ^{3/2}e^{-\mu (x^2+y^2)}\Bigg[\mu \left(3-\mu  \left(x^2+y^2\right)\right) \left(1-2 \mu  \left(x^2+y^2\right)\right)\nonumber\\
    &-\frac{\mu^2}{4}(kx^2+y^2) -\frac{1}{2}(x^4+y^4)-lx^2 y^2+\frac{3}{4}\mu y^3 +\frac{3}{4}n \mu x^2y\Bigg].
\end{align}

Using \eqref{e3.15}, we finally obtain the following coefficient
\begin{align}
    c_0=\int dx dy g(x,y)\Psi_0(x,y)= 2 \mu.
\end{align}

Like before, one has to construct normalized operators $K_1=\frac{1}{\sqrt{N_1}}K_1$, where the constant of normalization can be expressed as
\begin{align}
    N_1=(K_1|K_1)=(\mathcal{O}_1|\mathcal{O}_1)-4 \mu^2.
\end{align}

A straightforward computation reveals that
\begin{align}
  (\mathcal{O}_1|\mathcal{O}_1)=\int dx dy   g^2(x,y)=8 \mu ^2.
\end{align}

Combining them together, we find
\begin{align}
    N_1=4 \mu^2.
\end{align}

The next Krylov operator is given by
\begin{align}
\label{e3.24}
K_{2}=\mathcal{O}_{2}-c_{1}K_{1}-d_1 K_{0}.
\end{align}

Taking the inner product with $|K_0)$ and setting $(K_0|K_2)=0$ and $(K_0|K_1)=0$ , we find the value of the coefficient as
\begin{align}
\label{e3.25}
    d_1=(K_0|\mathcal{O}_2)=\int dx dy \Psi_0(x,y)h(x,y).
\end{align}

A straightforward computation of the integral \eqref{e3.25} yields
\begin{align}
    d_1=-\frac{1}{2 \mu }\Big[ (k-7) \mu ^3+l+3\Big].
\end{align}

On a similar note, we take the inner product with $|K_1)$ and setting $(K_1|K_2)=0$, we obtain the following coefficient
\begin{align}
\label{e3.27}
    c_1=\frac{1}{4\mu^2}\Big[(\mathcal{O}_1|\mathcal{O}_2)-c_0 d_1 \Big]=\frac{1}{4\mu^2}\int dx dy g(x,y)h(x,y)-\frac{c_0 d_1}{4 \mu^2}.
\end{align}

A straightforward evaluation of the integral \eqref{e3.27} reveals
\begin{align}
    c_1=\frac{1}{4 \mu ^2}\Big[ (k+25) \mu ^3+2 l+6\Big].
\end{align}

The normalization constant is given by
\begin{align}
    N_2=\frac{\mu ^4}{4} ((k-2) k+257) +\frac{\mu }{16}  (-128 l+27 n (n+2)-249)+\frac{(l^2+3)}{\mu ^2}.
\end{align}

In the following, we list normalized operators
\begin{align}
    &K_0=\mathcal{O}_0\\
    \label{e3.31}
  &K_1=\frac{1}{\sqrt{N_1}}\Big[ \mathcal{O}_1-c_0K_0\Big]\\
  &K_2=\frac{1}{\sqrt{N_2}}\Big[ \mathcal{O}_2-c_1K_1-d_1 K_0\Big].
  \label{e3.32}
\end{align}
\subsubsection{Lanczos coefficients and Krylov complexity}
With the above list of normalized operators, in the following we compute a couple of Lanczos coefficients, following the algorithm developed in \cite{Parker:2018yvk}. Like in the previous example of the Krylov state complexity, we will be mostly concerned with the early time ($t \sim 0$) growth of the operator ($\mathcal{O}(t)$). In other words, it is sufficient for us to compute the first few (off-diagonal) Lanczos coefficients ($b_n$) in the expansion. 

The (orthonormal) Krylov operators satisfy the Krylov chain condition \cite{Parker:2018yvk}
\begin{align}
\label{ee5.34}
    &A_{n+1}=\hat{\mathcal{L}}K_n-b_{n}K_{n-1}\\
    &A_{n+1}=b_{n+1}K_{n+1}.
\end{align}

Clearly, the diagonal entries are all zero by construction, that is,
\begin{align}
    a_n=L_{nn}=(K_n|\hat{\mathcal{L}}|K_n)=b_{n+1}(K_n|K_{n+1})+b_n(K_n|K_{n-1})=0.
\end{align}

The other set of (off-diagonal) Lanczos coefficients ($b_n$) are obtained by taking the inner product with $|K_{n-1})$, which yields the following \cite{Hashimoto:2023swv}
\begin{align}
    b_n=L_{n n-1} = (K_n| \hat{\mathcal{L}}|K_{n-1})
\end{align}
where we have used the cyclic property of trace of operators.

Notice that for $n=0$ we have $K_{-1}=0$, which yields $b_0=0$. The first non-zero Lanczos coefficient is obtained for $n=1$ (with $K_0=\mathcal{O}_0$)
\begin{align}
    b_1(\mu)=L_{10}=(K_1|\hat{\mathcal{L}}|K_0)=(K_1|[\mathcal{H},K_0])=(K_1|\mathcal{O}_1).
\end{align}

A further simplification reveals the following expression
\begin{align}
\label{e3.38}
    b_1(\mu)=\frac{1}{\sqrt{N}_1}\Big[(\mathcal{O}_1|\mathcal{O}_1)-2\mu (\mathcal{O}_0|\mathcal{O}_1)\Big]=2\mu.
\end{align}
Clearly, $b_1$ is a linear function of the mass deformation parameter $\mu$ as in the case of state complexity. However, the only difference appears to be in the constant pre-factor \cite{Roychowdhury:2026vzq}.

Next, we calculate the Lanczos coefficient $b_2$, which is given by 
\begin{align}
    b_2(\mu) =L_{21}=(K_2|\hat{\mathcal{L}}|K_1)=(K_2|[\mathcal{H},K_1]).
\end{align}

Using \eqref{e3.31}, one can further simplify this as
\begin{align}
    b_2 (\mu)&=\frac{1}{\sqrt{N}_1}(K_2|[\mathcal{H},\mathcal{O}_1])-\frac{2\mu}{\sqrt{N_1}}(K_2|[\mathcal{H},K_0])\nonumber\\
    &=\frac{1}{2\mu}(K_2|\mathcal{O}_2)-(K_2|\mathcal{O}_1).
\end{align}

Using \eqref{e3.32}, we obtain the following expression
\begin{align}
\label{e3.41}
    b_2(\mu)&=\frac{1}{2\mu \sqrt{N_2}}\Big[ (\mathcal{O}_2|\mathcal{O}_2)-\frac{c_1}{2\mu} (\mathcal{O}_1|\mathcal{O}_2)+\frac{c_0c_1}{2\mu} (K_0|\mathcal{O}_2)-d_1 (K_0|\mathcal{O}_2)\Big]\nonumber\\
    &-\frac{1}{\sqrt{N_2}}\Big[ (\mathcal{O}_2|\mathcal{O}_1)-\frac{c_1}{2\mu} (\mathcal{O}_1|\mathcal{O}_1)+\frac{c_0c_1}{2\mu} (K_0|\mathcal{O}_1)-d_1 (K_0|\mathcal{O}_1)\Big].
\end{align}

After performing the integrals in detail, and thus taking the large $\mu$ limit, we find
\begin{align}
    b_2(\mu)=\frac{(k (k+20)+341) \mu }{2 \sqrt{((k-2) k+257) }}-\frac{((k+17) (k+25)) }{8 \sqrt{((k-2) k+257) }}+\mathcal{O}(1/\mu^3).
\end{align}

The coefficients in the expansion \eqref{e3.12} satisfy the following equation \cite{Parker:2018yvk} 
\begin{align}
    \partial_t \varphi_n =b_n \varphi_{n-1}(t) -b_{n+1}\varphi_{n+1}(t).
\end{align}

Considering $n=0,1$ one arrives at the following set of equations
\begin{align}
\label{e3.44}
    & \partial_t \varphi_0 = -b_1 \varphi_1 (t)\\
    &\partial_t \varphi_1 =b_1 \varphi_0 (t) -b_2 \varphi_2 (t).
\end{align}

In order to compute $\varphi_1(t)$, we compare the expansions \eqref{e3.1} and \eqref{e3.12}
\begin{align}
    &\mathcal{O}(t)=\mathcal{O}_0+it \mathcal{O}_1-\frac{t^2}{2}\mathcal{O}_2+\cdots\\
    &\mathcal{O}(t)=\varphi_0 (t) K_0+i\varphi_1 (t)K_1-\varphi_2 (t)K_2+\cdots.
\end{align}

Taking the inner product with respect to $|K_1)$, we find
\begin{align}
\label{e3.48}
    \varphi_1(t)=2\mu t+\frac{it^2}{4 \mu}\Big[(\mathcal{O}_1|\mathcal{O}_2)-2\mu (K_0|\mathcal{O}_2) \Big].
\end{align}

A straightforward evaluation of the entities above in \eqref{e3.48} finally reveals
\begin{align}
   \varphi_1(t)&= 2 \mu  t+\frac{i t^2 }{4 \mu }\left((k+25) \mu ^3+2 l+6\right)
\end{align}
where we retain ourselves up to quadratic order in the expansion in time $t$.

From \eqref{e3.44}, this further produces the following expression for $\varphi_0 (t)$
\begin{align}
    \varphi_0(t)&=1-2 \mu ^2 t^2-\frac{i t^3}{6}  \left((k+25) \mu ^3+2 l+6\right)
\end{align}
where $\mu t $ is kept fixed in the limit $\mu \rightarrow \infty$ and $t \sim 0$. Notice that $\varphi_0(t=0)=1$, which is in accordance with the initial condition \cite{Parker:2018yvk}, \cite{Hashimoto:2023swv}, which sets $(\mathcal{O}(0)|\mathcal{O}(0))=1$.

Using the above information, we finally note down the function
\begin{align}
\label{e3.51}
    \varphi_2(t)=\kappa \varphi_0 (t)-\frac{1}{b_2}\partial_t \varphi_1(t)
\end{align}
where at LO in $\mu$, we notice that 
\begin{align}
\label{e3.52}
    \kappa = \frac{b_1}{b_2}=\frac{4\sqrt{((k-2) k+257) }}{(k (k+20)+341)}.
\end{align}

After some simplification, one finds that
\begin{align}
    &\varphi_2(t)= \kappa -\frac{i \kappa  t^3}{6}  \left((k+25) \mu ^3+2 l+6\right)-2 \kappa  \mu ^2 t^2\nonumber\\
    &-\frac{i \sqrt{(k-2) k+257} \left((k+25) \mu ^3 t-4 i \mu ^2+2 (l+3) t\right)}{(k (k+20)+341) \mu ^2}.
\end{align}

Taking into account a large $\mu$ expansion, one finds the following expression
\begin{align}
  \varphi_2(t)= -\frac{i \kappa }{6}  (k+25) \mu ^3 t^3-2\kappa  \mu ^2  t^2-\frac{i (k+25) \sqrt{(k-2) k+257} \mu  t}{k (k+20)+341} +\mathcal{O}(\mu^0).
\end{align}

Finally, the Krylov operator complexity \cite{Parker:2018yvk} can be expressed as
\begin{align}
\label{e3.55}
    &C(t)|_{t\sim 0}=\sum_{n}n |\varphi_n(t)|^2 =\zeta \mu^2 t^2+\mathcal{O}(t^4).
\end{align}

The coefficients above in \eqref{e3.55} can be expressed as
\begin{align}
    &\zeta(k)=2+\frac{(k-2) k (k+25)^2}{(k (k+20)+341)^2}+\frac{257 (k+25)^2}{(k (k+20)+341)^2}. 
\end{align}

Like in the case of the KSC, the leading behavior in $t$ is quadratic in nature. Notice that the coefficient $\zeta(k)$ of the LO term depends on the parameter $k$, which in turn depends on the FS model with which one is working. For example, for the pulsating fuzzy sphere model (PFS), one finds $\zeta=3.36586$. On the other hand, for integrable fuzzy sphere model (IFS) we have $k=1$, which yields $\zeta=3.32059$. Finally, in comparison with KSC, we notice that the first non-trivial correction due to massive deformation $\mu$ appears at same order in the time scale as in KSC, that is, it receives correction at order $\mu^2$ that scales as $t^2$.
\subsection{$N=3$ model: Coulomb potential}
The algorithm we follow here has already been mentioned above. The only change to be observed is in the state $\mathcal{O}_2$
\begin{align}
    \mathcal{O}_2=[\mathcal{H},\mathcal{O}_1]=\frac{1}{2}[p_x^2+p_y^2,\mathcal{O}_1]+[V(x,y),\mathcal{O}_1]
\end{align}
where $V(x,y)$ is given in \eqref{e2.2} and we use $\mu$ instead of $\bar{\mu}$.

After some algebra, one finds the following
\begin{align}
  &\mathcal{O}_2=  \frac{2 \sqrt{2}\mu^{3/2}}{\sqrt{\pi}}\Big[p^2_x+p^2_y-2 \mu (p_x x+p_y y)^2\Big]e^{-\mu (x^2+y^2)}  -\frac{8\sqrt{2}}{\sqrt{\pi}}\mu^{7/2}\Big(x^2+\frac{y^2}{4}\Big)e^{-\mu (x^2+y^2)}\nonumber\\
  &-\frac{16\sqrt{2}}{\sqrt{\pi}}\mu^{3/2}y^4 e^{-\mu (x^2+y^2)}+\frac{2\sqrt{2}}{\sqrt{\pi}}\mu^{3/2}\frac{\mathcal{Q}^2}{y^2}e^{-\mu (x^2+y^2)}-\frac{48\sqrt{2}}{\sqrt{\pi}}\mu^{3/2}x^2 y^2 e^{-\mu (x^2+y^2)}.
\end{align}

Like before, the action of $\mathcal{O}_2$ on the position eigen basis is defined as
\begin{align}
    &\mathcal{O}_2 \ket{x,y}=h(x,y)\ket{x,y}\\
    &h(x,y)=8\sqrt{\frac{2}{\pi }} \mu ^{3/2}e^{-\mu (x^2+y^2)}\Bigg[\mu \left(3-\mu  \left(x^2+y^2\right)\right) \left(1-2 \mu  \left(x^2+y^2\right)\right)\nonumber\\
    &-\mu^2\Big(x^2+\frac{y^2}{4}\Big) -2y^4 + \frac{\mathcal{Q}^2}{4 y^2}-6 x^2 y^2\Bigg].
\end{align}

Since the operators $\mathcal{O}_1$ and $K_0$ remain the same, therefore, the first nontrivial Krylov basis is the same as in \eqref{e3.31}. However, the correction appears for the operator $K_2$ \eqref{e3.24}, where one of the coefficients can be expressed as
\begin{align}
    d_1=\mu ^2 \left(\frac{3}{2}-8 \mathcal{Q}^2\right)-\frac{6}{\mu }.
\end{align}

The other coefficient \eqref{e3.27} can be expressed as
\begin{align}
    c_1=\mu  \left(\frac{29}{4}-4 \mathcal{Q}^2\right)+\frac{6}{\mu ^2}.
\end{align}

The normalized Krylov operator can be formally expressed as
\begin{align}
    K_2=\frac{1}{\sqrt{N_2}}\Big[ \mathcal{O}_2-c_1K_1-d_1 K_0\Big].
\end{align}

Following, a large $\mu$ expansion, the normalization constant can be expressed as
\begin{align}
    N_2=\mu ^4 \left(\frac{265}{4}-\frac{8}{3} \mathcal{Q}^2 \left(40 \mathcal{Q}^2+39\right)\right)+2 \mu  \left(16 \mathcal{Q}^2-57\right)+\frac{96}{\mu ^2}.
\end{align}

The first Lanczos coefficient $b_1(\mu)$ \eqref{e3.38} remains the same, since it depends on the states $\mathcal{O}_0$ and $\mathcal{O}_1$. The second non-vanishing Lanczos coefficient \eqref{e3.41} can be expressed as
\begin{align}
\label{e3.64}
    b_2(\mu)=\frac{\mu  \left(1311-32 \mathcal{Q}^2 \left(8 \mathcal{Q}^2+57\right)\right)}{2 \sqrt{3} \sqrt{ \left(795-32 \mathcal{Q}^2 \left(40 \mathcal{Q}^2+39\right)\right)}}+\mathcal{O}(1/\mu^3).
\end{align}

Calculating the Krylov operator growth requires the estimation of coefficients $\varphi_n(t)$. A straightforward calculation reveals the coefficient $\varphi_1(t)$ \eqref{e3.48} as
\begin{align}
\label{e3.68}
    \varphi_1(t)=2 \mu  t+\frac{i t^2 }{4 \mu }\left(\mu ^3 \left(29-16 \mathcal{Q}^2\right)+24\right)
\end{align}
which can be expanded in the large $\mu$ limit as
\begin{align}
    \varphi_1(t)=\frac{i \mu ^2 t^2}{4}  \left(29-16 \mathcal{Q}^2\right) +2 \mu  t+\mathcal{O}(1/\mu).
\end{align}

Using \eqref{e3.44}, one can calculate the coefficient $\varphi_0(t)$, which for the present case yields 
\begin{align}
\label{e3.67}
    \varphi_0(t)=1-2 \mu ^2 t^2+\frac{i t^3}{6}  \left(\mu ^3 \left(16 \mathcal{Q}^2-29\right)-24\right)
\end{align}
where we set the integration constant to one such that $\varphi_0(0)=1$.

Like before, one can expand \eqref{e3.67} in the large $\mu$ limit to yield
\begin{align}
    \varphi_0(t)=1-2 \mu ^2 t^2+\frac{i \mu ^3t^3}{6}  \left(16 \mathcal{Q}^2-29\right) +\mathcal{O}(t^3).
\end{align}

Finally, we use \eqref{e3.51} to calculate $\varphi_2(t)$, which after a straightforward calculation yields 
\begin{align}
\label{e3.72}
    &\varphi_2(t)=\kappa +\frac{i \kappa  t^3}{6}  \left(\mu ^3 \left(16 \mathcal{Q}^2-29\right)-24\right)-2 \kappa  \mu ^2 t^2\nonumber\\
    &+\frac{\sqrt{2385-96 \mathcal{Q}^2 \left(40 \mathcal{Q}^2+39\right)} \left(4 \mu ^2+i t \left(\mu ^3 \left(29-16 \mathcal{Q}^2\right)+24\right)\right)}{\mu ^2 \left(32 \mathcal{Q}^2 \left(8 \mathcal{Q}^2+57\right)-1311\right)}
\end{align}

Considering a large $\mu$ expansion, one finds that
\begin{align}
    &\varphi_2(t)=-\frac{2 i \mu ^3t^3 \left(16 \mathcal{Q}^2-29\right) \sqrt{2385-96 \mathcal{Q}^2 \left(40 \mathcal{Q}^2+39\right)} }{96 \mathcal{Q}^2 \left(8 \mathcal{Q}^2+57\right)-3933}\nonumber\\
    &+\frac{8 \mu ^2 t^2\sqrt{2385-96 \mathcal{Q}^2 \left(40 \mathcal{Q}^2+39\right)} }{32 \mathcal{Q}^2 \left(8 \mathcal{Q}^2+57\right)-1311}+\mathcal{O}(\mu t).
\end{align}

Here $\kappa$ is defined as before in \eqref{e3.52}, while $b_2$ is given in \eqref{e3.64}. Using \eqref{e3.68} and \eqref{e3.72}, we obtain the KOC in the large $\mu$ limit as
\begin{align}
\label{e3.73}
    C(t)|_{t \sim 0}=\zeta \mu^2 t^2+\mathcal{O}(t^4).
\end{align}

\begin{figure}
    \centering
    \includegraphics[width=0.7\linewidth]{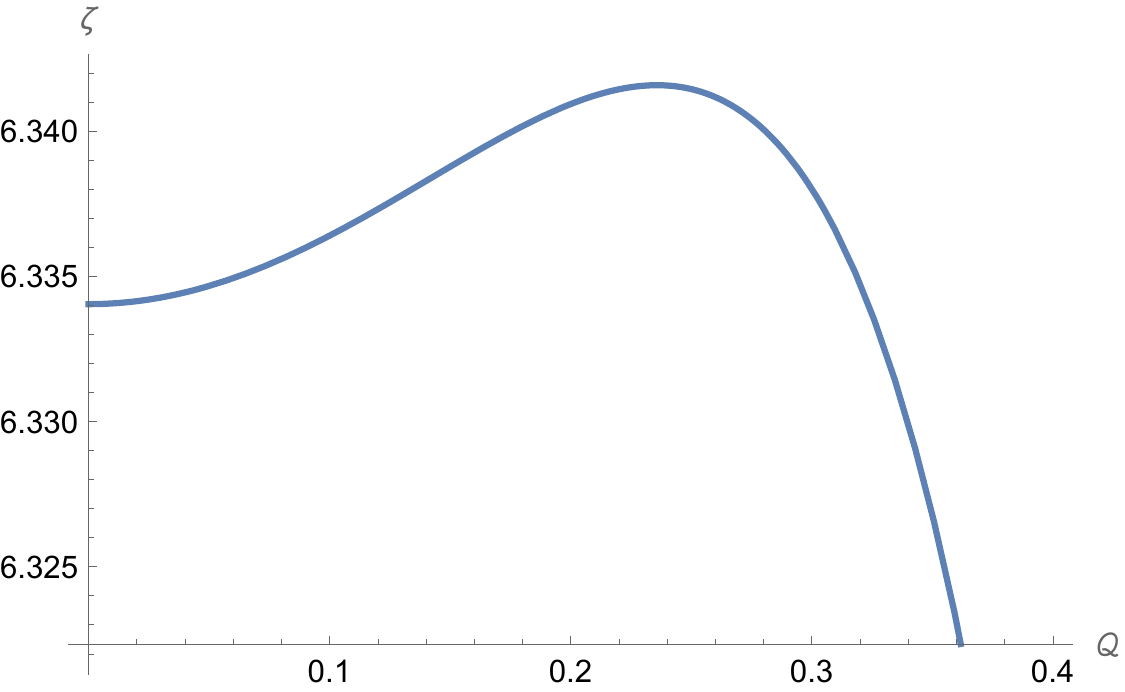}
    \caption{Plot the LO coefficient $\zeta$ with the Coulomb charge $\mathcal{Q}$.}
    \label{figzeta}
\end{figure}

The coefficients above in \eqref{e3.73} can be expressed as
\begin{align}
    &\zeta (\mathcal{Q})=\frac{2}{\left(1311-32 \mathcal{Q}^2 \left(8 \mathcal{Q}^2+57\right)\right)^2}\Bigg[ 5443227-64 \mathcal{Q}^2 \Big(233235 \nonumber\\
    &+3328 \mathcal{Q}^4 \left(4 \mathcal{Q}^2-21\right)-96360 \mathcal{Q}^2\Big)\Bigg]. 
\end{align}

Notice that the leading order coefficient is model dependent, that is, is depends on the Coulomb charge $\mathcal{Q}$. The above analysis is valid for $\mathcal{Q}\ll 1$. A closer analysis reveals that the coefficient $\zeta$ initially increases with $\mathcal{Q}$, resulting in a larger effect on the complexity of the operator growth. However, beyond some critical threshold $\mathcal{Q}=\mathcal{Q}_c$, it starts to fall, resulting in a decrease in the operator complexity, see Fig.\ref{figzeta}. This feature is similar to that observed in the case of Krylov complexity of states (Fig.\ref{fig1}).
\section{Summary and conclusions}
\label{sec4}
In summary, we discuss the Krylov state and operator complexity considering various systematic reductions of the BMN Plane Wave Matrix Model. These include $N=2,3,4$ representations of the matrices in the matrix model. In each case, the model reduces to a Hamiltonian dynamics that enables us to carry out a systematic analysis of complexity, either for states or operators. Our analysis reveals a universality class among the Lanczos coefficients, considering different reduction ansatz, in the limit of strong mass deformation, $\mu \rightarrow \infty$. We show that the generic form of the Lanczos coefficients at LO in large mass deformation could be schematically expressed as $a_n=\alpha_n \mu$ and $b_n = \beta_n \mu$, where the coefficients $\alpha_n$ and $\beta_n$ depend on the particular reduction ansatz along with the type of complexity sought (see Table \ref{table}). In other words, the Lanczos coefficients vary linearly with the mass deformation parameter $\mu$. On the other hand, the Krylov state (KSC) and Krylov operator complexity (KOC) receive corrections at quadratic order in $\mu$, that is, $C(t)=\zeta \mu^2 t^2$. We notice that the massive correction in both complexities appears in order $t^2$. The coefficient $\zeta>0$ is model dependent and controls the early time growth of the complexity.

Before we conclude, we mention below a number of interesting future directions that may be pursued in the future. They include the following points.

$\bullet$ It would be nice to extend the above analysis for the \emph{intermediate} coupling $\mu$. A comparison of results in the intermediate and weak coupling would be really interesting.

$\bullet$ One has to map the results of the present paper and the dual gravity calculations. In particular, it would be nice to find the dual counterpart of the Krylov state complexity and classify different notions of complexity from the holographic perspective. An immediate step forward would be to be able to compute the Lanczos coefficients from a gravity perspective. Lanczos coefficients can be obtained from moments that should be computable from two point function or the geodesic length on the gravity counterpart.

$\bullet$ On the matrix model side, there are various limits of mass deformation that can be considered. It would be nice to identify how these limits are realized on the gravity side.

We hope to address some of these issues in the near future.

\paragraph {Acknowledgements :}
  The author thanks Carlos Nunez for discussion. The author also acknowledges the Mathematical Research Impact Centric Support (MATRICS) grant no. (MTR/2023/000005) received from ANRF, India. \\ 


\begin{thebibliography}{99}

\bibitem{Parker:2018yvk}
D.~E.~Parker, X.~Cao, A.~Avdoshkin, T.~Scaffidi and E.~Altman,
``A Universal Operator Growth Hypothesis,''
Phys. Rev. X \textbf{9}, no.4, 041017 (2019)
doi:10.1103/PhysRevX.9.041017
[arXiv:1812.08657 [cond-mat.stat-mech]].

\bibitem{Balasubramanian:2022tpr}
V.~Balasubramanian, P.~Caputa, J.~M.~Magan and Q.~Wu,
``Quantum chaos and the complexity of Krylov of states,''
Phys. Rev. D \textbf{106}, no.4, 046007 (2022)
doi:10.1103/PhysRevD.106.046007
[arXiv:2202.06957 [hep-th]].

\bibitem{Balasubramanian:2022dnj}
V.~Balasubramanian, J.~M.~Magan and Q.~Wu,
``Tridiagonalizing random matrices,''
Phys. Rev. D \textbf{107}, no.12, 126001 (2023)
doi:10.1103/PhysRevD.107.126001
[arXiv:2208.08452 [hep-th]].

\bibitem{Dymarsky:2021bjq}
A.~Dymarsky and M.~Smolkin,
``Krylov complexity in conformal field theory,''
Phys. Rev. D \textbf{104}, no.8, L081702 (2021)
doi:10.1103/PhysRevD.104.L081702
[arXiv:2104.09514 [hep-th]].

\bibitem{Avdoshkin:2022xuw}
A.~Avdoshkin, A.~Dymarsky and M.~Smolkin,
``Krylov complexity in quantum field theory, and beyond,''
JHEP \textbf{06}, 066 (2024)
doi:10.1007/JHEP06(2024)066
[arXiv:2212.14429 [hep-th]].

\bibitem{Kundu:2023hbk}
A.~Kundu, V.~Malvimat and R.~Sinha,
``State dependence of Krylov complexity in 2d CFTs,''
JHEP \textbf{09}, 011 (2023)
doi:10.1007/JHEP09(2023)011
[arXiv:2303.03426 [hep-th]].

\bibitem{Alishahiha:2022anw}
M.~Alishahiha and S.~Banerjee,
``A universal approach to Krylov state and operator complexities,''
SciPost Phys. \textbf{15}, no.3, 080 (2023)
doi:10.21468/SciPostPhys.15.3.080
[arXiv:2212.10583 [hep-th]].

\bibitem{Caputa:2025dep}
P.~Caputa and G.~Di Giulio,
``Local quenches from a Krylov perspective,''
JHEP \textbf{07}, 164 (2025)
doi:10.1007/JHEP07(2025)164
[arXiv:2502.19485 [hep-th]].

\bibitem{Caputa:2025ozd}
P.~Caputa, G.~Di Giulio and T.~Q.~Loc,
``Symmetry-Resolved Krylov Complexity,''
[arXiv:2509.12992 [hep-th]].

\bibitem{Caputa:2024vrn}
P.~Caputa, H.~S.~Jeong, S.~Liu, J.~F.~Pedraza and L.~C.~Qu,
``Krylov complexity of density matrix operators,''
JHEP \textbf{05}, 337 (2024)
doi:10.1007/JHEP05(2024)337
[arXiv:2402.09522 [hep-th]].

\bibitem{Hashimoto:2023swv}
K.~Hashimoto, K.~Murata, N.~Tanahashi and R.~Watanabe,
``Krylov complexity and chaos in quantum mechanics,''
JHEP \textbf{11}, 040 (2023)
doi:10.1007/JHEP11(2023)040
[arXiv:2305.16669 [hep-th]].

\bibitem{Barbon:2019wsy}
J.~L.~F.~Barb{\'o}n, E.~Rabinovici, R.~Shir and R.~Sinha,
``On The Evolution Of Operator Complexity Beyond Scrambling,''
JHEP \textbf{10}, 264 (2019)
doi:10.1007/JHEP10(2019)264
[arXiv:1907.05393 [hep-th]].

\bibitem{Baggioli:2024wbz}
M.~Baggioli, K.~B.~Huh, H.~S.~Jeong, K.~Y.~Kim and J.~F.~Pedraza,
``Krylov complexity as an order parameter for quantum chaotic-integrable transitions,''
Phys. Rev. Res. \textbf{7}, no.2, 023028 (2025)
doi:10.1103/PhysRevResearch.7.023028
[arXiv:2407.17054 [hep-th]].

\bibitem{Alishahiha:2024vbf}
M.~Alishahiha, S.~Banerjee and M.~J.~Vasli,
``Krylov complexity as a probe for chaos,''
Eur. Phys. J. C \textbf{85}, no.7, 749 (2025)
doi:10.1140/epjc/s10052-025-14451-z
[arXiv:2408.10194 [hep-th]].

\bibitem{Bhattacharjee:2024yxj}
B.~Bhattacharjee and P.~Nandy,
``Krylov fractality and complexity in generic random matrix ensembles,''
Phys. Rev. B \textbf{111}, no.6, L060202 (2025)
doi:10.1103/PhysRevB.111.L060202
[arXiv:2407.07399 [quant-ph]].

\bibitem{Caputa:2024sux}
P.~Caputa, B.~Chen, R.~W.~McDonald, J.~Sim{\'o}n and B.~Strittmatter,
``Spread complexity rate as proper momentum,''
Phys. Rev. D \textbf{113}, no.4, L041901 (2026)
doi:10.1103/7zs8-9zpg
[arXiv:2410.23334 [hep-th]].

\bibitem{Fatemiabhari:2025poq}
A.~Fatemiabhari, H.~Nastase, C.~Nunez and D.~Roychowdhury,
``Holographic Krylov complexity for conformal quiver gauge theories,''
Nucl. Phys. B \textbf{1025}, 117402 (2026)
doi:10.1016/j.nuclphysb.2026.117402
[arXiv:2512.14812 [hep-th]].

\bibitem{Fatemiabhari:2026rob}
A.~Fatemiabhari, C.~Nunez and R.~T.~Santamaria,
``Complexity and Operator Growth in Holographic 6d SCFTs,''
[arXiv:2603.10106 [hep-th]].

\bibitem{Fatemiabhari:2026goj}
A.~Fatemiabhari and C.~Nunez,
``Krylov Complexity, Confinement and Universality,''
[arXiv:2602.17757 [hep-th]].

\bibitem{Fatemiabhari:2025usn}
A.~Fatemiabhari, H.~Nastase, C.~Nunez and D.~Roychowdhury,
``Holographic Operator complexity in confining gauge theories,''
[arXiv:2511.22717 [hep-th]].

\bibitem{Fatemiabhari:2025cyy}
A.~Fatemiabhari, H.~Nastase and D.~Roychowdhury,
``Holographic Operator complexity in ${\cal N}=4$ SYM,''
[arXiv:2511.19286 [hep-th]].

\bibitem{Roychowdhury:2026eds}
D.~Roychowdhury,
``Holographic Krylov complexity for Yang-Baxter deformed supergravity backgrounds,''
[arXiv:2601.06555 [hep-th]].

\bibitem{Nastase:2026lhz}
H.~Nastase, C.~Nunez and D.~Roychowdhury,
``Holographic Krylov Complexity for Charged, Composite and Extended Probes,''
[arXiv:2604.07432 [hep-th]].

\bibitem{Jeong:2026iac}
H.~S.~Jeong,
``Krylov Subspace Dynamics as Near-Horizon AdS$_2$ Holography,''
[arXiv:2602.11627 [hep-th]].

\bibitem{Heller:2024ldz}
M.~P.~Heller, J.~Papalini and T.~Schuhmann,
``Krylov Krylov complexity as holographic complexity beyond JT gravity,''
[arXiv:2412.17785 [hep-th]].

\bibitem{Heller:2025ddj}
M.~P.~Heller, F.~Ori, J.~Papalini, T.~Schuhmann and M.~T.~Wang,
``De Sitter holographic complexity from Krylov complexity in DSSYK,''
[arXiv:2510.13986 [hep-th]].

\bibitem{Fu:2025kkh}
Y.~Fu, H.~S.~Jeong, K.~Y.~Kim and J.~F.~Pedraza,
``Toward Krylov-based holography in double-scaled SYK,''
[arXiv:2510.22658 [hep-th]].

\bibitem{Roychowdhury:2026sgg}
D.~Roychowdhury,
``Krylov complexity for Lin-Maldacena geometries and their holographic duals,''
[arXiv:2604.16977 [hep-th]].

\bibitem{Ambrosini:2024sre}
M.~Ambrosini, E.~Rabinovici, A.~S{\'a}nchez-Garrido, R.~Shir and J.~Sonner,
``Operator K-complexity in DSSYK: Krylov complexity equals bulk length,''
JHEP \textbf{08}, 059 (2025)
doi:10.1007/JHEP08(2025)059
[arXiv:2412.15318 [hep-th]].

\bibitem{Rabinovici:2023yex}
E.~Rabinovici, A.~S{\'a}nchez-Garrido, R.~Shir and J.~Sonner,
``A bulk manifestation of Krylov complexity,''
JHEP \textbf{08}, 213 (2023)
doi:10.1007/JHEP08(2023)213
[arXiv:2305.04355 [hep-th]].

\bibitem{Erdmenger:2022lov}
J.~Erdmenger, A.~L.~Weigel, M.~Gerbershagen and M.~P.~Heller,
``From complexity geometry to holographic spacetime,''
Phys. Rev. D \textbf{108}, no.10, 106020 (2023)
doi:10.1103/PhysRevD.108.106020
[arXiv:2212.00043 [hep-th]].

\bibitem{Kar:2021nbm}
A.~Kar, L.~Lamprou, M.~Rozali and J.~Sully,
``Random matrix theory for complexity growth and black hole interiors,''
JHEP \textbf{01}, 016 (2022)
doi:10.1007/JHEP01(2022)016
[arXiv:2106.02046 [hep-th]].

\bibitem{Berenstein:2002jq}
D.~E.~Berenstein, J.~M.~Maldacena and H.~S.~Nastase,
``Strings in flat space and pp waves from N=4 superYang-Mills,''
JHEP \textbf{04}, 013 (2002)
doi:10.1088/1126-6708/2002/04/013
[arXiv:hep-th/0202021 [hep-th]].

\bibitem{Asano:2015eha}
Y.~Asano, D.~Kawai and K.~Yoshida,
``Chaos in the BMN matrix model,''
JHEP \textbf{06}, 191 (2015)
doi:10.1007/JHEP06(2015)191
[arXiv:1503.04594 [hep-th]].

\bibitem{Dasgupta:2002hx}
K.~Dasgupta, M.~M.~Sheikh-Jabbari and M.~Van Raamsdonk,
``Matrix perturbation theory for M theory on a PP wave,''
JHEP \textbf{05}, 056 (2002)
doi:10.1088/1126-6708/2002/05/056
[arXiv:hep-th/0205185 [hep-th]].

\bibitem{Sugiyama:2002rs}
K.~Sugiyama and K.~Yoshida,
``Supermembrane on the PP wave background,''
Nucl. Phys. B \textbf{644}, 113-127 (2002)
doi:10.1016/S0550-3213(02)00794-0
[arXiv:hep-th/0206070 [hep-th]].

\bibitem{Banks:1996vh}
T.~Banks, W.~Fischler, S.~H.~Shenker and L.~Susskind,
``M theory as a matrix model: A conjecture,''
Phys. Rev. D \textbf{55}, 5112-5128 (1997)
doi:10.1201/9781482268737-37
[arXiv:hep-th/9610043 [hep-th]].

\bibitem{Lin:2004nb}
H.~Lin, O.~Lunin and J.~M.~Maldacena,
``Bubbling AdS space and 1/2 BPS geometries,''
JHEP \textbf{10}, 025 (2004)
doi:10.1088/1126-6708/2004/10/025
[arXiv:hep-th/0409174 [hep-th]].

\bibitem{Lin:2005nh}
H.~Lin and J.~M.~Maldacena,
``Fivebranes from gauge theory,''
Phys. Rev. D \textbf{74}, 084014 (2006)
doi:10.1103/PhysRevD.74.084014
[arXiv:hep-th/0509235 [hep-th]].

\bibitem{Ling:2006up}
H.~Ling, A.~R.~Mohazab, H.~H.~Shieh, G.~van Anders and M.~Van Raamsdonk,
``Little string theory from a double-scaled matrix model,''
JHEP \textbf{10}, 018 (2006)
doi:10.1088/1126-6708/2006/10/018
[arXiv:hep-th/0606014 [hep-th]].

\bibitem{Asano:2014vba}
Y.~Asano, G.~Ishiki, T.~Okada and S.~Shimasaki,
``Emergent bubbling geometries in the plane wave matrix model,''
JHEP \textbf{05}, 075 (2014)
doi:10.1007/JHEP05(2014)075
[arXiv:1401.5079 [hep-th]].

\bibitem{Asano:2012zt}
Y.~Asano, G.~Ishiki, T.~Okada and S.~Shimasaki,
``Exact results for perturbative partition functions of theories with SU(2|4) symmetry,''
JHEP \textbf{02}, 148 (2013)
doi:10.1007/JHEP02(2013)148
[arXiv:1211.0364 [hep-th]].

\bibitem{Polchinski:2000uf}
J.~Polchinski and M.~J.~Strassler,
``The String dual of a confining four-dimensional gauge theory,''
[arXiv:hep-th/0003136 [hep-th]].

\bibitem{Lin:2004kw}
H.~Lin,
``The Supergravity dual of the BMN matrix model,''
JHEP \textbf{12}, 001 (2004)
doi:10.1088/1126-6708/2004/12/001
[arXiv:hep-th/0407250 [hep-th]].

\bibitem{Myers:1999ps}
R.~C.~Myers,
``Dielectric branes,''
JHEP \textbf{12}, 022 (1999)
doi:10.1088/1126-6708/1999/12/022
[arXiv:hep-th/9910053 [hep-th]].

\bibitem{Amore:2024ihm}
P.~Amore, L.~A.~Pando Zayas, J.~F.~Pedraza, N.~Quiroz and C.~A.~Terrero-Escalante,
``Fuzzy spheres in stringy matrix models: quantifying chaos in a mixed phase space,''
JHEP \textbf{06}, 031 (2025)
doi:10.1007/JHEP06(2025)031
[arXiv:2407.07259 [hep-th]].

\bibitem{Roychowdhury:2026vzq}
D.~Roychowdhury,
``Krylov state complexity for BMN matrix model,''
[arXiv:2605.10786 [hep-th]].

\bibitem{Huh:2024ytz}
K.~B.~Huh, H.~S.~Jeong, L.~A.~Pando Zayas and J.~F.~Pedraza,
``Krylov complexity in mixed phase space,''
Phys. Rev. D \textbf{111}, no.12, L121902 (2025)
doi:10.1103/gmy7-dn7l
[arXiv:2412.04963 [hep-th]].

\bibitem{Arnlind:2003nh}
J.~Arnlind and J.~Hoppe,
``Classical solutions in the BMN matrix model,''
[arXiv:hep-th/0312166 [hep-th]].

\bibitem{saha}
R. Sahadevan and M. Lakshmanan, Invariance and integrability: Henon-Heiles and two coupled quartic anharmonic oscillator systems, J. Phys. A: Math. Gen. 19 (1986) L949.


\end{thebibliography}
\end{document}